\documentclass[twocolumn]{aastex631}

\usepackage{placeins}

\usepackage{gensymb}
\usepackage{comment}
\usepackage{float}

\newcommand{\CHECK}[1]{{#1}}
\newcommand{\cytwo}[1]{#1}

\reportnum{DES-2025-0921}
\reportnum{FERMILAB-PUB-25-0714-LDRD-PPD}

\begin{document}

\title{Ultra-Faint Milky Way Satellites Discovered in Carina, Phoenix, and Telescopium  \\ with DELVE Data Release 3} 

\correspondingauthor{Chin Yi Tan}
\email{chinyi@uchicago.edu}
\author[0000-0003-0478-0473]{C.~Y.~Tan}
\affiliation{Kavli Institute for Cosmological Physics, University of Chicago, Chicago, IL 60637, USA}
\affiliation{Department of Physics, University of Chicago, Chicago, IL 60637, USA}
\affiliation{NSF-Simons AI Institute for the Sky (SkAI),172 E. Chestnut St., Chicago, IL 60611, USA}
\author[0000-0003-1697-7062]{W.~Cerny}
\altaffiliation{Contributed Equally}
\affiliation{Department of Astronomy, Yale University, New Haven, CT 06520, USA}

\author[0000-0002-6021-8760]{A.~B.~Pace}
\altaffiliation{Contributed Equally}
\affiliation{Department of Astronomy, University of Virginia, 530 McCormick Road, Charlottesville, VA 22904, USA}
\author[0009-0001-1133-5047]{J.~A.~Sharp}
\affiliation{Kavli Institute for Cosmological Physics, University of Chicago, Chicago, IL 60637, USA}
\affiliation{Department of Astronomy and Astrophysics, University of Chicago, Chicago, IL 60637, USA}

\author[0009-0008-0959-0162]{K.~Overdeck}
\affiliation{Kavli Institute for Cosmological Physics, University of Chicago, Chicago, IL 60637, USA}
\affiliation{Department of Astronomy and Astrophysics, University of Chicago, Chicago, IL 60637, USA}
\affiliation{NSF-Simons AI Institute for the Sky (SkAI),172 E. Chestnut St., Chicago, IL 60611, USA}

\author[0000-0001-8251-933X]{A.~Drlica-Wagner}
\affiliation{Fermi National Accelerator Laboratory, P.O.\ Box 500, Batavia, IL 60510, USA}
\affiliation{Kavli Institute for Cosmological Physics, University of Chicago, Chicago, IL 60637, USA}
\affiliation{Department of Astronomy and Astrophysics, University of Chicago, Chicago, IL 60637, USA}
\affiliation{NSF-Simons AI Institute for the Sky (SkAI),172 E. Chestnut St., Chicago, IL 60611, USA}

\author[0000-0002-4733-4994]{J.~D.~Simon}
\affiliation{Observatories of the Carnegie Institution for Science, 813 Santa Barbara St., Pasadena, CA 91101, USA}
 \author[0000-0001-9649-4815]{B.~Mutlu-Pakdil}
 \affiliation{Department of Physics and Astronomy, Dartmouth College, Hanover, NH 03755, USA}
\author[0000-0003-4102-380X]{D.~J.~Sand}
\affiliation{Department of Astronomy/Steward Observatory, 933 North Cherry Avenue, Room N204, Tucson, AZ 85721-0065, USA}

\author[0009-0004-5519-0929]{A.~M.~Senkevich}
\affiliation{Department of Physics, University of Surrey, Guildford GU2 7XH, UK}

\author[0000-0002-8448-5505 ]{D.~Erkal}
\affiliation{Department of Physics, University of Surrey, Guildford GU2 7XH, UK}

\author[0000-0001-6957-1627]{P.~S.~Ferguson}
\affiliation{DiRAC Institute, Department of Astronomy, University of Washington, 3910 15th Ave NE, Seattle, WA, 98195, USA}

\author[0000-0002-7822-0658]{F.~Sobreira}
\affiliation{Laboratório Interinstitucional de e-Astronomia- LIneA, Rua Gal. José Cristino 77, Rio de Janeiro, RJ- 20921-400, Brazil}
\affiliation{Instituto de Fisica Gleb Wataghin, Universidade Estadual de Campinas, 13083-859, Campinas, SP, Brazil}

 \author[0000-0001-9649-8103]{K.~R.~Atzberger}
 \affiliation{Department of Astronomy, University of Virginia, 530 McCormick Road, Charlottesville, VA 22904, USA}

 \author[0000-0002-3936-9628]{J.~L.~Carlin}
 \affiliation{Rubin Observatory/AURA, 950 North Cherry Avenue, Tucson, AZ, 85719, USA}

 \author[0000-0002-7155-679X]{A.~Chiti}
 \affiliation{Department of Astronomy and Astrophysics, University of Chicago, Chicago, IL 60637, USA}
 \affiliation{Kavli Institute for Cosmological Physics, University of Chicago, Chicago, IL 60637, USA}

\author[0000-0002-1763-4128]{D.~Crnojevi\'c}
\affiliation{Department of Physics \& Astronomy, University of Tampa, 401 West Kennedy Boulevard, Tampa, FL 33606, USA}

\author[0000-0002-4863-8842]{A.~P.~Ji}
\affiliation{Kavli Institute for Cosmological Physics, University of Chicago, Chicago, IL 60637, USA}
\affiliation{Department of Astronomy and Astrophysics, University of Chicago, Chicago, IL 60637, USA}

\author[0000-0001-6421-0953]{L.~C.~Johnson}
\affiliation{Center for Interdisciplinary Exploration and Research in Astrophysics (CIERA) and Department of Physics and Astronomy, Northwestern University, 1800 Sherman Ave, Evanston, IL 60201 USA}

\author[0000-0002-9110-6163]
{T.~S.~Li}
\affiliation{Department of Astronomy and Astrophysics, University of Toronto, 50 St. George Street, Toronto ON, M5S 3H4, Canada}

 \author[0000-0002-9269-8287]{G.~Limberg}
 \affiliation{Kavli Institute for Cosmological Physics, University of Chicago, Chicago, IL 60637, USA}

 \author[0000-0002-9144-7726]{C.~E.~Mart\'inez-V\'azquez}
 \affiliation{International Gemini Observatory/NSF NOIRLab, 670 N. A'ohoku Place, Hilo, Hawai'i, 96720, USA}
 \author[0000-0003-0105-9576]{G.~E.~Medina}
 \affiliation{Department of Astronomy and Astrophysics, University of Toronto, 50 St. George Street, Toronto ON, M5S 3H4, Canada}

\author[0000-0003-4479-1265]{V.~M.~Placco}
 \affiliation{NSF NOIRLab, 950 N. Cherry Ave., Tucson, AZ 85719, USA}

 \author[0000-0001-5805-5766]{A.~H.~Riley}
  \affiliation{Lund Observatory, Division of Astrophysics, Department of Physics, Lund University, SE-221 00 Lund, Sweden}
 \affiliation{Institute for Computational Cosmology, Department of Physics, Durham University, South Road, Durham DH1 3LE, UK}

\author[0000-0002-9599-310X]{E.~J.~Tollerud}
 \affiliation{Space Telescope Science Institute, 3700 San Martin Drive, Baltimore, MD 21218, USA}

\author[0000-0003-4341-6172]{A.~K.~Vivas}
\affiliation{Cerro Tololo Inter-American Observatory/NSF NOIRLab, Casilla 603, La Serena, Chile}


\author[0000-0003-1587-3931]{T.~M.~C.~Abbott}
\affiliation{Cerro Tololo Inter-American Observatory/NSF NOIRLab, Casilla 603, La Serena, Chile}

\author[0000-0001-5679-6747]{M.~Aguena}
\affiliation{INAF-Osservatorio Astronomico di Trieste, via G. B. Tiepolo 11, I-34143 Trieste, Italy}
\affiliation{Laborat\'orio Interinstitucional de e-Astronomia - LIneA, Av. Pastor Martin Luther King Jr, 126 Del Castilho, Nova Am\'erica Offices, Torre 3000/sala 817 CEP: 20765-000, Brazil}

\author[0000-0002-7394-9466]{O.~Alves}
\affiliation{Department of Physics, University of Michigan, Ann Arbor, MI 48109, USA}

\author[[0000-0002-2562-8537]{D.~Bacon}
\affiliation{Institute of Cosmology and Gravitation, University of Portsmouth, Portsmouth, PO1 3FX, UK}

\author[0000-0002-4900-805X]{S.~Bocquet}
\affiliation{University Observatory, LMU Faculty of Physics, Scheinerstr. 1, 81679 Munich, Germany}

\author[0000-0002-8458-5047]{D.~Brooks}
\affiliation{Department of Physics \& Astronomy, University College London, Gower Street, London, WC1E 6BT, UK}

\author[0000-0003-1866-1950]{D.~L.~Burke}
\affiliation{Kavli Institute for Particle Astrophysics \& Cosmology, P. O. Box 2450, Stanford University, Stanford, CA 94305, USA}
\affiliation{SLAC National Accelerator Laboratory, Menlo Park, CA 94025, USA}

\author[0000-0002-7436-3950]{R.~Camilleri}
\affiliation{School of Mathematics and Physics, University of Queensland,  Brisbane, QLD 4072, Australia}

 \author[0000-0002-3690-105X]{J.~A.~Carballo-Bello}
 \affiliation{Instituto de Alta Investigaci\'on, Universidad de Tarapac\'a, Casilla 7D, Arica, Chile}

\author[0000-0003-3044-5150]{A.~Carnero~Rosell}
\affiliation{Instituto de Astrofisica de Canarias, E-38205 La Laguna, Tenerife, Spain}
\affiliation{Laborat\'orio Interinstitucional de e-Astronomia - LIneA, Av. Pastor Martin Luther King Jr, 126 Del Castilho, Nova Am\'erica Offices, Torre 3000/sala 817 CEP: 20765-000, Brazil}
\affiliation{Universidad de La Laguna, Dpto. Astrofísica, E-38206 La Laguna, Tenerife, Spain}

\author[0000-0002-3130-0204]{J.~Carretero}
\affiliation{Institut de F\'{\i}sica d'Altes Energies (IFAE), The Barcelona Institute of Science and Technology, Campus UAB, 08193 Bellaterra (Barcelona) Spain}

 \author[0000-0001-8670-4495, gname='Ting-Yun', sname='Cheng']{T.-Y.~Cheng}
\affiliation{Kapteyn Astronomical Institute, University of Groningen, Landleven 12 (Kapteynborg, 5419), 9747 AD Groningen, The Netherlands}

 \author[0000-0003-1680-1884]{Y.~Choi}
 \affiliation{NSF NOIRLab, 950 N. Cherry Ave., Tucson, AZ 85719, USA}

\author[0000-0002-7731-277X]{L.~N.~da Costa}
\affiliation{Laborat\'orio Interinstitucional de e-Astronomia - LIneA, Av. Pastor Martin Luther King Jr, 126 Del Castilho, Nova Am\'erica Offices, Torre 3000/sala 817 CEP: 20765-000, Brazil}

\author[0000-0002-7131-7684]{M.~E.~da Silva Pereira}
\affiliation{Hamburger Sternwarte, Universit\"{a}t Hamburg, Gojenbergsweg 112, 21029 Hamburg, Germany}

\author[0000-0002-4213-8783]{T.~M.~Davis}
\affiliation{School of Mathematics and Physics, University of Queensland,  Brisbane, QLD 4072, Australia}

\author[0000-0001-8318-6813]{J.~De~Vicente}
\affiliation{Centro de Investigaciones Energ\'eticas, Medioambientales y Tecnol\'ogicas (CIEMAT), Madrid, Spain}

\author[0000-0002-0466-3288]{S.~Desai}
\affiliation{Department of Physics, IIT Hyderabad, Kandi, Telangana 502285, India}

\author[0000-0002-6397-4457]{P.~Doel}
\affiliation{Department of Physics \& Astronomy, University College London, Gower Street, London, WC1E 6BT, UK}

\author[0000-0002-3745-2882]{S.~Everett}
\affiliation{California Institute of Technology, 1200 East California Blvd, MC 249-17, Pasadena, CA 91125, USA}

\author[0000-0002-2367-5049]{B.~Flaugher}
\affiliation{Fermi National Accelerator Laboratory, P. O. Box 500, Batavia, IL 60510, USA}

\author[0000-0003-4079-3263]{J.~Frieman}
\affiliation{Department of Astronomy and Astrophysics, University of Chicago, Chicago, IL 60637, USA}
\affiliation{Fermi National Accelerator Laboratory, P. O. Box 500, Batavia, IL 60510, USA}
\affiliation{Kavli Institute for Cosmological Physics, University of Chicago, Chicago, IL 60637, USA}

\author[0000-0002-9370-8360]{J.~Garc\'ia-Bellido}
\affiliation{Instituto de Fisica Teorica UAM/CSIC, Universidad Autonoma de Madrid, 28049 Madrid, Spain}

\author[0000-0003-3270-7644]{D.~Gruen}
\affiliation{University Observatory, LMU Faculty of Physics, Scheinerstr. 1, 81679 Munich, Germany}

\author[0000-0003-0825-0517]{G.~Gutierrez}
\affiliation{Fermi National Accelerator Laboratory, P. O. Box 500, Batavia, IL 60510, USA}

\author[0000-0003-2071-9349]{S.~R.~Hinton}
\affiliation{School of Mathematics and Physics, University of Queensland,  Brisbane, QLD 4072, Australia}

\author[0000-0002-9369-4157]{D.~L.~Hollowood}
\affiliation{Santa Cruz Institute for Particle Physics, Santa Cruz, CA 95064, USA}

\author[0000-0001-5160-4486]{D.~J.~James}
\affiliation{Center for Astrophysics $\vert$ Harvard \& Smithsonian, 60 Garden Street, Cambridge, MA 02138, USA}

\author[0000-0003-0120-0808]{K.~Kuehn}
\affiliation{Australian Astronomical Optics, Macquarie University, North Ryde, NSW 2113, Australia}
\affiliation{Lowell Observatory, 1400 Mars Hill Rd, Flagstaff, AZ 86001, USA}

\author[0000-0002-8289-740X]{S.~Lee}
\affiliation{Jet Propulsion Laboratory, California Institute of Technology, 4800 Oak Grove Dr., Pasadena, CA 91109, USA}

\author[0000-0003-0710-9474]{J.~L.~Marshall}
\affiliation{George P. and Cynthia Woods Mitchell Institute for Fundamental Physics and Astronomy, and Department of Physics and Astronomy, Texas A\&M University, College Station, TX 77843,  USA}

\author[0000-0001-9497-7266]{J. Mena-Fern{\'a}ndez}
\affiliation{Universit\'e Grenoble Alpes, CNRS, LPSC-IN2P3, 38000 Grenoble, France}

\author[0000-0002-1372-2534]{F.~Menanteau}
\affiliation{Center for Astrophysical Surveys, National Center for Supercomputing Applications, 1205 West Clark St., Urbana, IL 61801, USA}
\affiliation{Department of Astronomy, University of Illinois at Urbana-Champaign, 1002 W. Green Street, Urbana, IL 61801, USA}

\author[0000-0002-6610-4836]{R.~Miquel}
\affiliation{Instituci\'o Catalana de Recerca i Estudis Avan\c{c}ats, E-08010 Barcelona, Spain}
\affiliation{Institut de F\'{\i}sica d'Altes Energies (IFAE), The Barcelona Institute of Science and Technology, Campus UAB, 08193 Bellaterra (Barcelona) Spain}

\author[0000-0001-6145-5859]{J.~Myles}
\affiliation{Department of Astrophysical Sciences, Princeton University, Peyton Hall, Princeton, NJ 08544, USA}

\author[0000-0001-9438-5228]{M.~Navabi}
\affiliation{Department of Physics, University of Surrey, Guildford GU2 7XH, UK}

\author[0000-0002-1793-3689]{D.~L.~Nidever}
\affiliation{Department of Physics, Montana State University, P.O. Box 173840, Bozeman, MT 59717-3840}

\author[0000-0002-8282-469X]{N.~E.~D.~No\"el}
\affiliation{Department of Physics, University of Surrey, Guildford GU2 7XH, UK}

\author[0000-0003-2120-1154, gname='Ricardo', sname='Ogando']{R.~L.~C.~Ogando}
\affiliation{Centro de Tecnologia da Informa\c{c}\~ao Renato Archer, Campinas, SP - 13069-901, Brazil }
\affiliation{Observat\'orio Nacional, Rua Gal. Jos\'e Cristino 77, Rio de Janeiro, RJ - 20921-400, Brazil}

\author[0000-0002-2598-0514]{A.~A.~Plazas~Malag\'on}
\affiliation{Kavli Institute for Particle Astrophysics \& Cosmology, P. O. Box 2450, Stanford University, Stanford, CA 94305, USA}
\affiliation{SLAC National Accelerator Laboratory, Menlo Park, CA 94025, USA}

\author[0000-0002-2762-2024]{A.~Porredon}
\affiliation{Centro de Investigaciones Energ\'eticas, Medioambientales y Tecnol\'ogicas (CIEMAT), Madrid, Spain}
\affiliation{Ruhr University Bochum, Faculty of Physics and Astronomy, Astronomical Institute, German Centre for Cosmological Lensing, 44780 Bochum, Germany}

\author[0000-0001-7147-8843]{S.~Samuroff}
\affiliation{Department of Physics, Northeastern University, Boston, MA 02115, USA}
\affiliation{Institut de F\'{\i}sica d'Altes Energies (IFAE), The Barcelona Institute of Science and Technology, Campus UAB, 08193 Bellaterra (Barcelona) Spain}

\author[0000-0002-9646-8198]{E.~Sanchez}
\affiliation{Centro de Investigaciones Energ\'eticas, Medioambientales y Tecnol\'ogicas (CIEMAT), Madrid, Spain}

\author[0000-0003-3054-7907]{D.~Sanchez Cid}
\affiliation{Centro de Investigaciones Energ\'eticas, Medioambientales y Tecnol\'ogicas (CIEMAT), Madrid, Spain}
\affiliation{Physik-Institut, University of Zürich, Winterthurerstrasse 190, CH-8057 Zürich, Switzerland}

\author[0000-0002-1831-1953]{I.~Sevilla-Noarbe}
\affiliation{Centro de Investigaciones Energ\'eticas, Medioambientales y Tecnol\'ogicas (CIEMAT), Madrid, Spain}

\author[0000-0002-3321-1432]{M.~Smith}
\affiliation{Physics Department, Lancaster University, Lancaster, LA1 4YB, UK}

\author[0000-0003-1479-3059]{G.~S.~Stringfellow}
\affiliation{Center for Astrophysics and Space Astronomy, University of Colorado, 389 UCB, Boulder, CO 80309-0389, USA}

\author[0000-0002-7047-9358]{E.~Suchyta}
\affiliation{Computer Science and Mathematics Division, Oak Ridge National Laboratory, Oak Ridge, TN 37831}

\author[0000-0002-1488-8552]{M.~E.~C.~Swanson}
\affiliation{Center for Astrophysical Surveys, National Center for Supercomputing Applications, 1205 West Clark St., Urbana, IL 61801, USA}

\author{V.~Vikram}
\affiliation{Department of Physics, Central University of Kerala, 93VR+RWF, CUK Rd, Kerala 671316, India}

\author[0000-0002-7123-8943, gname='Alistair', sname='Walker']{A.~R.~Walker}
\affiliation{Cerro Tololo Inter-American Observatory/NSF NOIRLab, Casilla 603, La Serena, Chile}

\author[0000-0001-6455-9135]{A.~Zenteno}
\affiliation{Cerro Tololo Inter-American Observatory/NSF NOIRLab, Casilla 603, La Serena, Chile}

\collaboration{79}{(DELVE and DES Collaborations)}

\begin{abstract}
We report the discovery of three Milky Way satellite candidates: Carina IV, Phoenix III, and DELVE 7, in the third data release of the DECam Local Volume Exploration survey (DELVE). The candidate systems were identified by cross-matching results from two independent search algorithms. All three are extremely faint systems composed of old, metal-poor stellar populations ($\tau \gtrsim 10$ Gyr, [Fe/H] $ \lesssim -1.4$).  Carina IV (\CHECK{$M_V = -2.8;\ r_{1/2} = 40\,{\rm pc}$)} and Phoenix III \CHECK{($M_V = -1.2;\ r_{1/2} = 19\,{\rm pc}$)} have half-light radii that are consistent with the known population of dwarf galaxies, while DELVE 7 \CHECK{($M_V = 1.2;\ r_{1/2} = 2\,{\rm pc}$)} is very compact and seems more likely to be a star cluster, though its nature remains ambiguous without spectroscopic followup. The \textit{Gaia} proper motions of stars in Carina IV \CHECK{($M_\star = 2250^{+1180}_{-830}\,{\rm M_\odot}$)} indicate that it is unlikely to be associated with the LMC, while DECam CaHK photometry confirms that its member stars are metal-poor. Phoenix III \CHECK{($M_\star = 520^{+660}_{-290}\,{\rm M_\odot}$)} is the faintest known satellite  in the extreme outer stellar halo \CHECK{($D_{\rm GC} > 100$\,kpc)}, while DELVE~7 (\CHECK{$M_\star = 60^{+120}_{-40}\,{\rm M_\odot}$}) is the faintest known satellite with $D_{\rm GC} > 20$ kpc.
\end{abstract}
 
\section{INTRODUCTION}
\label{sec:intro}

Ultra-faint dwarf galaxies (UFDs) are the faintest, oldest, and most dark-matter-dominated stellar systems known. These galaxies have absolute magnitudes fainter than $M_V=-7.7$ ($L=10^5$~L$_\odot$) \citep{Simon:2019}, and their extremely low luminosity has made them especially powerful probes of the physical processes that regulate galaxy evolution, and the nature of dark matter (e.g., \citealt{Bullock:2017, Simon:2019} and references therein). These systems occupy the extreme limit of galaxy formation and allow us to probe how atomic hydrogen cooling and reionization set the minimum mass of a dark matter halo that is capable of hosting a galaxy \citep{2000ApJ...539..517B, Okamoto:2008, 2024MNRAS.529.3387A, 2025MNRAS.540.1107S}. On the other hand, the high dark-matter fractions and relatively simple internal dynamics of UFDs let us test dark-matter physics on extremely small spatial scales (e.g., tens of parsecs; \citealt{Bullock:2017}). For example, the census of low-mass dark matter subhalos around the Milky Way (as traced by its satellite galaxies) already places lower limits on the dark matter particle mass \citep{Kennedy:2014, Jethwa:2018, 2021JCAP...08..062N, 2021MNRAS.506.5848E, MW2, 2025MNRAS.tmp.1192N, Tan:2025b}, and UFDs are prime targets for indirect searches for dark matter annihilation or decay \citep[e.g.,][]{Ackermann:2015zua, Geringer-Sameth:2015, 2024PhRvD.109f3024M, 2025ApJ...978L..43C}. Nevertheless, because these galaxies are so faint, most of the known population exists as satellites of the Milky Way or of other nearby hosts in the Local Volume \citep[D $<$ 12 Mpc; e.g.,][]{2022ApJ...933...47C, Pace:2024}.

The first ``ultra-faint'' Milky Way satellites were discovered in data from the Sloan Digital Sky Survey \citep[SDSS; e.g.,][]{Willman:2005a, Willman:2005b, Belokurov07, Belokurov08, Grillmair:2009}. Subsequent wide-field surveys, such as the Pan-STARRS-1 $3\pi$ survey \citep[{PS1:}][]{Chambers:2016, 2015ApJ...802L..18L, 2015ApJ...813...44L}, the Dark Energy Survey \citep[DES:][]{DES:2016, Bechtol:2015, Koposov15a, Kim:2015b}, and the Hyper Suprime-Cam Strategic Survey Program \citep[HSC-SSP;][]{Aihara:2018, 2019PASJ...71...94H, Homma:2024} pushed the observational frontier to even lower luminosities, increasing the total to $\sim$65 spectroscopically confirmed Milky Way satellite galaxies and likely galaxy candidates \citep{Pace:2024}. The discovery of extremely faint satellites has accelerated due to increasingly sensitive observations; however, many galaxies are still predicted to remain undetected \citep[e.g.,][]{MW2, Tsiane:2025, DELVECensus1}.

The faintest confirmed dwarf galaxy, Tucana V, has an absolute magnitude of $M_V \sim -1.1$ \citep{Simon:2020, Hansen:2024}, corresponding to a stellar mass of $M_\star \sim 500$ M$_\odot$. Even fainter and more compact satellites have been reported, including DELVE~1 \citep{Mau:2020}, Eridanus~III \citep{Simon:2024}, and Ursa Major~III/UNIONS~1 \citep{Smith:2024} (see \citealt{Pace:2024} for examples). However, the nature of these systems is unclear because the locus of dwarf galaxies in the size--luminosity plane overlaps that of disrupting star clusters at $M_V \gtrsim -3.5$ and velocity dispersions measurements of these faint systems are at the limit of current observational capabilities. Recent theoretical work suggests that extremely low-luminosity galaxies could exist \cytwo{\citep[e.g.,][]{Manwadkar:2022, Errani:2020, Errani:2024b, Ahvazi:2025}}, offering an exciting new regime to test models of galaxy formation and dark matter physics.
At the same time, it is likely that some of these systems are star clusters that originated within low-mass galaxies that were subsequently accreted onto the Milky Way.
Thus, the nature of ultra-faint compact satellites remains actively debated.

In order to better address questions about the nature of the least-luminous stellar systems in the halo of the Milky Way, the DECam Local Volume Exploration survey \citep[DELVE;][]{Drlica-Wagner:2021} is imaging the high-Galactic-latitude southern sky to search for faint Milky Way satellites. In a companion paper, we present the largest systematic census of Milky Way satellites to date, designed to quantify the detection efficiencies of our surveys \citep{DELVECensus1}. That work contains a systematic search for Milky Way satellites accompanied by a detailed characterization of the survey selection function. That analysis necessarily set a high threshold on detection significance to yield a 100\% pure sample of real systems. Here, we lower the detection significance threshold and identify compelling new systems.

In this paper, we report the discovery of three new ultra-faint satellites in the constellations Carina, Phoenix, and Telescopium.
Based on morphological analysis, two of the candidates have physical sizes consistent with the population of Milky Way satellite galaxies. Following established conventions, we designate them Carina IV\footnote{After Carina \citep{Cannon:1977}, Carina II, and Carina III \citep{Torrealba:2018}.} and Phoenix III\footnote{After Phoenix \citep{Schuster:1976} and Phoenix II \citep{Bechtol:2015, Koposov15a}.}. The third system is more compact and resembles an outer-halo star cluster, thus we follow convention and designate it DELVE 7\footnote{After DELVE~1--6 \citep{Mau:2020, Cerny:2021b, Cerny:2023c, Cerny:2023b}.}. 

This paper is organized as follows. In Section~\ref{sec:data} we describe the DELVE data used for our search. In Section~\ref{sec:search}, we describe our automated search for Milky Way satellite candidates. In Section~\ref{sec:measurement}, we characterize the properties of the new systems using data from DELVE and follow-up imaging with Magellan/IMACS for DELVE 7. For the brightest system, Carina IV, we present follow-up observations and analyses that confirm its nature as a Milky Way satellite in Section~\ref{sec:followup}. We briefly discuss the new systems in Section~\ref{sec:discussion} and summarize in Section~\ref{sec:summary}.

\section{DELVE Data}
\label{sec:data}

The DELVE program (PropID: 2019A-0305) seeks to discover and characterize dwarf galaxy satellites around the Milky Way, Magellanic Clouds, and isolated Magellanic analogs in the Local Group \citep{Drlica-Wagner:2021, Drlica-Wagner:2022}. To achieve this, DELVE uses the Dark Energy Camera \citep[DECam;][]{flaugher_2015_decam} on the Víctor M. Blanco 4-meter Telescope at the NSF Cerro Tololo Inter-American Observatory (CTIO) in Chile to image the high-Galactic-latitude southern sky in the $g$, $r$, $i$, and $z$ bands. To date, DELVE has been allocated more than 150 nights  to augment public archival DECam data, including data from DES \citep{DES:2016}, the DECam Legacy Survey \citep[DECaLS;][]{Dey:2019} and the DECam eROSITA Survey \citep[DeROSITAS;][]{Zenteno:2025}.  DELVE DR3 (\citealt{Tan:2025}; Drlica-Wagner et al.\ in prep.)\footnote{\url{https://datalab.noirlab.edu/data/delve}} includes source catalogs produced from coadded images processed uniformly using the DES Data Management pipeline \citep[DESDM;][]{Morganson:2018}, with image detrending and coaddition following DES DR2 \citep{DES:2021}.

Consistency between the DES and DELVE processing allows us to merge the DES Y6 Gold catalog \citep{Bechtol:2025} with the new DELVE catalogs to cover an unmasked sky area of \CHECK{$\sim$20,000\,deg$^2$}. After masking regions of high stellar density (where the {\it Gaia} DR2 source density exceeds 8 stars arcmin$^{-2}$ for $G<21$; \citealt{Gaia:2018}) and regions of high interstellar extinction ($E(B-V) > 0.2$), the final search footprint spans \CHECK{$\sim$17,000\,deg$^2$} (see Figure 2 of \citealt{Tan:2025b}). Within the $\sim$5,000\,deg$^2$ DES footprint, the imaging data are deeper and more homogeneous, reaching median 10$\sigma$ point-source depths of \CHECK{$g\sim24.7$, $r\sim24.4$, $i\sim23.8$} with a standard deviation of 0.2 mag across the three bands. Outside the DES footprint, the DELVE coverage is somewhat shallower and less uniform, with median depths of \CHECK{$g\sim24.2$, $r\sim23.7$, $i\sim23.2$} and a larger variation of \CHECK{$\sim0.5$} mag across the three bands, reflecting the heterogeneous field coverage of the survey \citep{Tan:2025b}.

To identify Milky Way satellites, we search for overdensities consistent with old, metal-poor stellar populations. The DELVE DR3 catalogs include stars belonging to the target Milky Way satellites, foreground Milky Way stars, and distant
background galaxies. To obtain a high-purity stellar sample from the catalogs, we select objects with \texttt{EXT\_XGB} $\leq 1$. The flag \texttt{EXT\_XGB} classifies objects into different morphological classes  using the \texttt{XGBoost} algorithm \citep{Chen:2016, Bechtol:2025}, with integer classes ranging from 0 (high-confidence stars) to 4 (high-confidence galaxies). \cytwo{As an additional check, we examine galaxy density maps around the candidates, constructed using non-stellar sources with $\texttt{EXT\_XGB}>2$, to ensure that the detected overdensities arise from genuine stellar systems rather than from misclassified galaxies.} We also exclude objects with 
\texttt{GOLD\_FLAGS} $> 0$ to remove sources with possible data-processing issues or unphysical measurements. Details of both flags can be found in \citet{Bechtol:2025}.

Our photometric measurements of the stars come from point-spread function (PSF) model photometry using the \texttt{PSF\_MAG\_APER\_8} values \citep{Bechtol:2025}. These measurements come from a simultaneous, multi-epoch fit using the individual PSF models for each image at the object location. The flux is normalized to the \texttt{MAG\_APER\_8} system to match the convention of \texttt{SourceExtractor} \citep{Bertin:1996, Bertin:2002}. We account for extinction due to interstellar dust by using dereddened measurements with the \texttt{\_CORRECTED} suffix, calculated by applying the interstellar extinction correction $A_b = R_b \times E(B - V)$ where $R_g = 3.186$, $R_r = 2.140$, $R_i = 1.569$, and $R_z = 1.196$ \citep{DES:2021}. The $E(B - V)$ values are obtained from the \citet{Schlegel1998} reddening maps with the calibration adjustment suggested by \citet{Schlafly:2011}. By convention, we denote extinction-corrected magnitudes with the subscript “0”.

\begin{figure*}
    \centering
    \includegraphics[width= \linewidth]{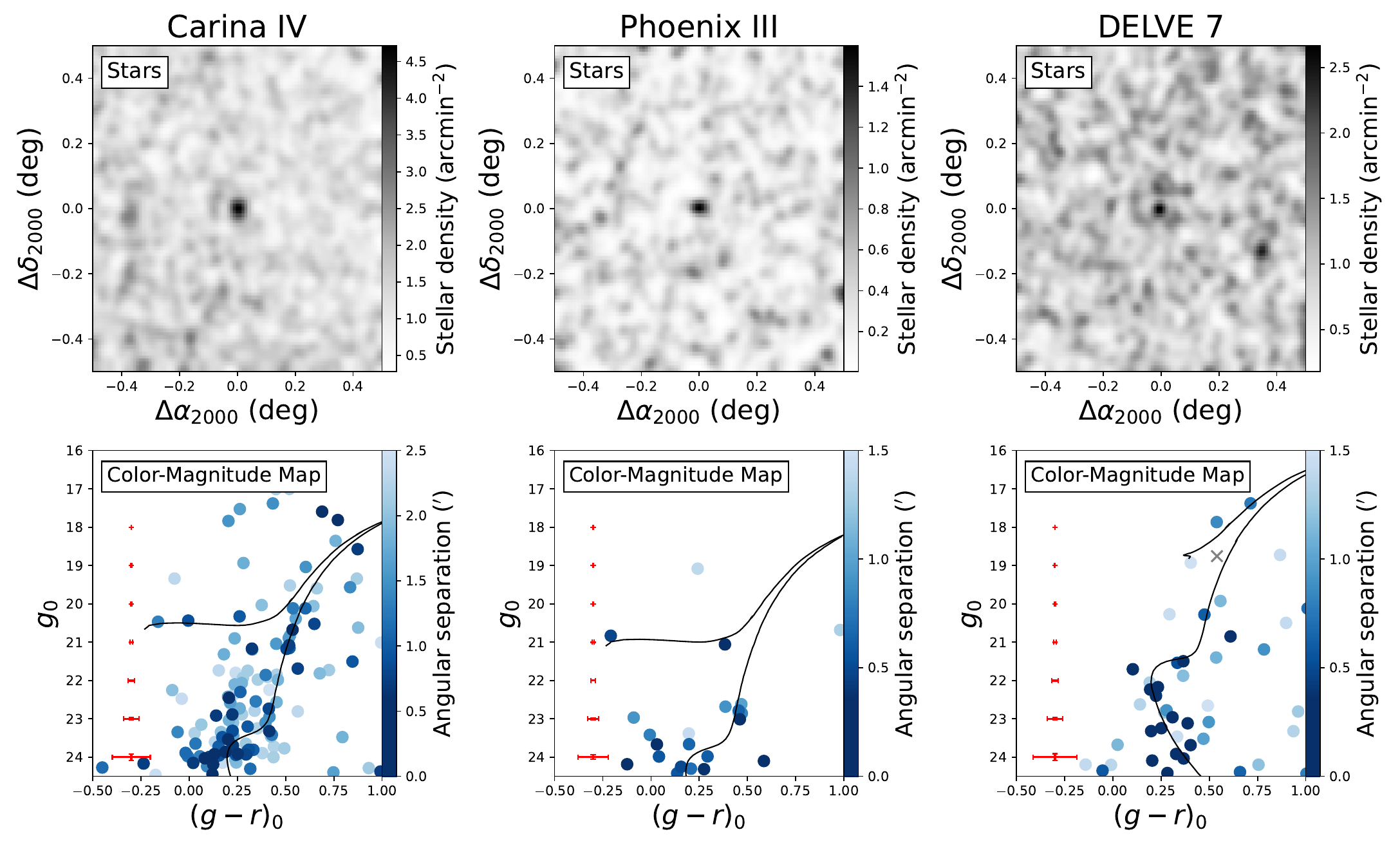}
    \includegraphics[width=0.97 \textwidth, clip, trim={0 0 0.5cm 1cm}]{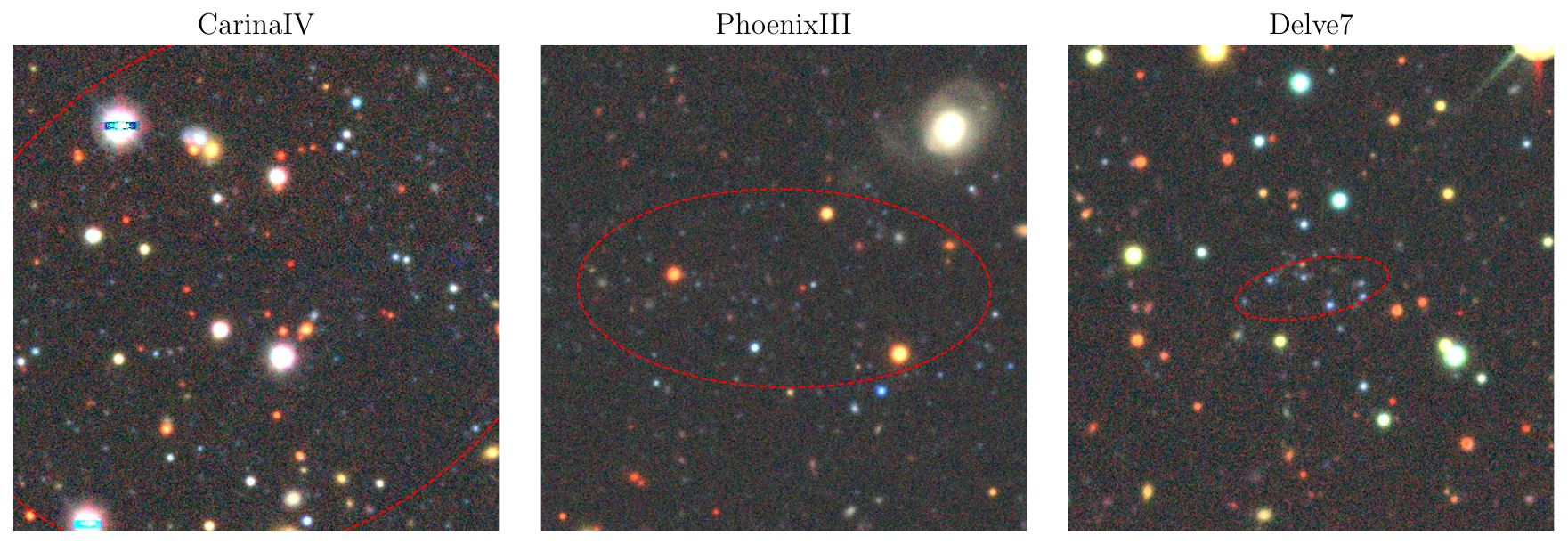}

    \caption{Diagnostic plots for each of the three newly-discovered satellites, closely matching those used to identify the systems in our search. (Top) \cytwo{Isochrone filtered} stellar density maps for a $1\degree\times1\degree$ degree region centered on each candidate, smoothed with a \CHECK{$0\farcm5$} Gaussian kernel.  (Middle) Color--magnitude diagrams for stars within $2\farcm5$ of Carina IV and $1\farcm5$ of Phoenix III and DELVE 7, centered on each candidate system. Stars are colored by separation from each system's centroid, with darker colors representing smaller separations. Also overlaid is the best-fit old, metal-poor \texttt{PARSEC} isochrone, with parameters taken from the best-fit \texttt{ugali} results in Table \ref{table:params}.  \cytwo{The median photometric uncertainties for stars around each satellite, as a function of $g$-band magnitude, are indicated by red error bars.  The “x” marker indicates a confirmed non–member star of DELVE~7 with a \textit{Gaia} parallax of $\varpi = 0.71 \pm 0.15$~mas, clearly placing it in the foreground of the system.} (Bottom) False-color coadded images ($\sim 2\arcmin \times 2\arcmin$, $\sim 0.03\degree \times 0.03\degree$) of Carina~IV, Phoenix~III, and  DELVE~7 from the Legacy Surveys Sky Viewer DR10. Each cutout has been artificially brightened by a factor of $1.8\times$ to highlight faint sources. \cytwo{We also overlay  red ellipses on the coadded images demarcating the half-light radius, $r_h$, of the satellites. \label{fig:diagnostic_plot}}}
\end{figure*}

\section{Dwarf Search Methods}
\label{sec:search}

We use two algorithms to search for new Milky Way satellites in
the DELVE DR3 data: the isochrone matched filter algorithm  \texttt{simple}\footnote{\url{https://github.com/DarkEnergySurvey/simple}} and the likelihood-based 
Ultra-faint GAlaxy LIkelihood toolkit \texttt{ugali}\footnote{\url{https://github.com/DarkEnergySurvey/ugali}} \citep{Bechtol:2015, MWCensus1}.  These two algorithms show similar performance when detecting real systems; however, false positives identified by one algorithm are often not detected by the other. Therefore, by requiring a candidate to be independently detected by both algorithms, we significantly reduce the number of false positives (see Figure 4 of \citealt{DELVECensus1}).

The first search algorithm, \texttt{simple}, has been used to discover at least \CHECK{29} ultra-faint Milky Way satellites using DECam data \citep{Bechtol:2015, Drlica-Wagner:2015, Mau:2020, Cerny:2021,Cerny:2021b,Cerny:2023c,Cerny:2023,Cerny:2023b, Cerny:2024, Tan:2025}. The \texttt{simple} algorithm first selects stars in color--magnitude space that are consistent with an old ($\tau = 12\,\mathrm{Gyr}$), metal-poor ($Z = 0.0001$, ${\rm [Fe/H]} \sim -2.2$) PARSEC isochrone \citep{Bressan:2012}. This is done by selecting stars that are consistent with $\Delta(g-r)_0 < \sqrt{0.1^2 + \sigma_g^2 + \sigma_r^2}, $
where $\sigma_g$ and $\sigma_r$ are the $g$- and $r$-band magnitude uncertainties of the individual stars, and the constant term $0.1$ accounts for additional modeling and systematic uncertainties.  Candidates are then identified by flagging regions that show high densities of isochrone-filtered stars. The \texttt{simple} detection significance of a candidate, SIG, is calculated by comparing the number of observed stars within a circular aperture centered on the candidate, $N_{\rm obs}$, to the surrounding background density field:
\begin{equation}
    {\rm SIG}_{gr} = {\rm ISF}_\mathcal{N} \left[ {\rm SF}_{\mathcal{P}(\lambda=N_b)} (N_{\rm obs}) \right],
\end{equation}
where ${\rm SF}_{\mathcal{P}(\lambda=N_b)}$ is the survival function of a Poisson distribution with mean, $N_b$, estimated from a background annulus and ${\rm ISF}_\mathcal{N}$ is the inverse survival function of a Gaussian normal distribution, $\mathcal{N}(\mu=0, \sigma=1)$.  When searching for candidates, we iterate over circular apertures with radii from $0\farcm6$ to $18\arcmin$ in steps of $0\farcm6$, and over distance moduli in the range $16.0 \leq (m-M)_0 \leq 23.0$\,mag in steps of $0.5$\,mag (corresponding to heliocentric distances $10\,{\rm kpc} < D_\odot < 400\,{\rm kpc}$).

The second search algorithm, \texttt{ugali}, employs a maximum-likelihood approach to identify galaxy candidates. Specifically, it compares the likelihood of a model that includes a Milky Way satellite galaxy against a null model consisting only of a uniform field-star background.  \cytwo{The uniform field-star background population is determined empirically using stars from a circular annulus surrounding each candidate ($0\fdg5 < r < 2\fdg0$), where they are binned in color–magnitude space.} The likelihood is constructed by evaluating the membership probability of stars within $r < 0\fdg5$ of the candidate system. Each star’s probability of belonging to a putative dwarf galaxy is determined based on its spatial position, photometry, and color. Full details of the likelihood construction are provided in \citet{Bechtol:2015} and  Appendix C of \citet{MWCensus1}.

We model the spatial distribution of stars in satellite galaxies with a radially symmetric Plummer profile \citep{Plummer:1911}. For the color--magnitude distribution, we build probability density functions in the \{$g_0$, $(g - r)_0$\} plane using old, metal-poor {\texttt{PARSEC v1.2S}} isochrones \citep{Bressan:2012, Chen:2014, Tang:2014, Chen:2015}, weighted by a \citet{Chabrier:2001} initial mass function (IMF). For our search, we adopt a composite isochrone obtained by summing four equally weighted isochrones with metallicities, $Z_{\rm phot} = \{0.0001, 0.0002\}$ (${\rm [Fe/H]} \approx \{-2.2, -1.9\}$), and ages, $\tau = \{10\,{\rm Gyr}, 12\,{\rm Gyr}\}$.

We evaluate the likelihood across a grid of \texttt{HEALPix}\footnote{http://healpix.sf.net} pixels (\texttt{nside} = 4096; spatial resolution $0\farcm08$). At each pixel, we vary the half-light radius, $r_h = \{1\farcm2, 4\farcm2, 9\farcm0\}$ and the distance modulus $16.0 < (m - M)_0 < 23.0$\,mag in 0.5\,mag steps, selecting the parameter set that maximizes the likelihood. The \texttt{ugali}  detection significance is quantified as the likelihood ratio between the satellite model and the null model, \mbox{$   \rm{TS}_{gr} = 2 ( \log\mathcal{L}_{\rm galaxy} - log\mathcal{L}_{\rm null})$}. We identify candidate systems by locating isolated peaks (i.e., contiguous regions) in the likelihood maps with $\mathrm{TS}_{gr} > 10$.

\begin{figure*}
    \centering
    \includegraphics[width=\linewidth]{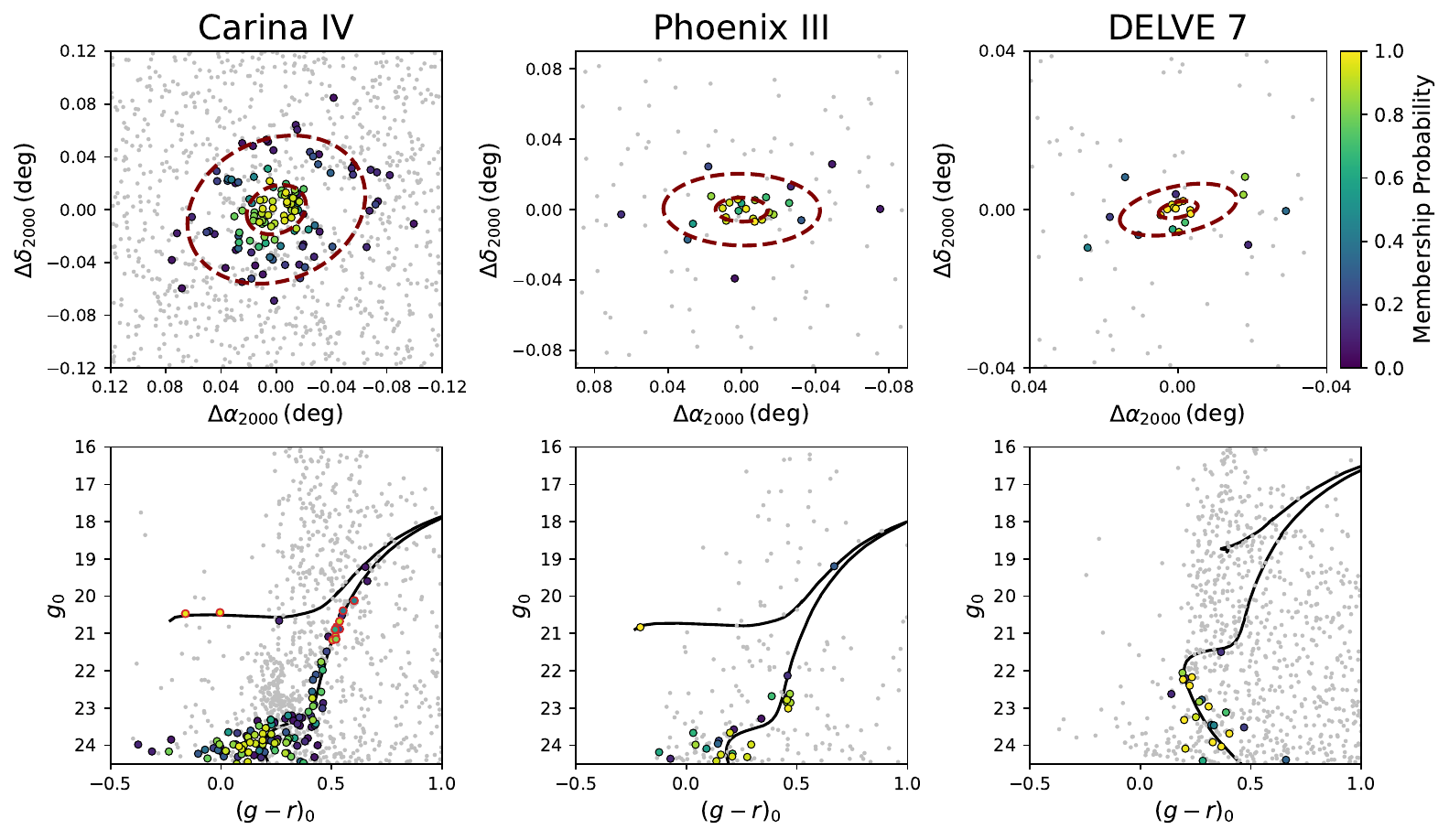}\
    
    \caption{\label{fig:ugali} The spatial distributions (top) and color--magnitude diagrams (bottom) for likely member stars in Carina~IV (left), Phoenix~III (center), and DELVE~7 (right). Stars are colored by their membership probabilities derived as part of our \texttt{ugali} characterization analysis, which incorporates the spatial, photometric, and color information of each star.  Stars  with a lower \texttt{ugali} membership probability ($p_{\rm ugali} < 0.05$) are shown in grey. The top panels also show ellipses representing the \texttt{ugali} best-fit half-light radius, $r_h$, and three times this half-light radius, $3r_h$, for each system. \cytwo{For Carina IV, we highlight with red outlines the stars in the color–magnitude diagram that have both a high \texttt{ugali} membership probability ($p_{\rm ugali} > 0.20$) and a \textit{Gaia} detection. These stars are discussed further in Section \ref{sec:gaia} and Appendix \ref{appendix:carina4_stars}.} Carina~IV is the best-populated of the three systems, featuring a clear main-sequence turn off (MSTO), red giant branch (RGB), and blue horizontal branch (BHB). Phoenix~III is at a similar distance, but it is fainter with only a handful of likely member stars. DELVE~7 is substantially fainter, featuring no highly-probable RGB members and is potentially only comprised of main-sequence stars. This lack of RGB stars is consistent with other faint Milky Way satellites at $M_V > 0$ (e.g., DELVE~5, Kim~3, and Ursa Major~III/UNIONS~1). }
\end{figure*}

Running our search algorithms on the DELVE DR3 data, we obtain tens of thousands of ``hotspots''  (i.e., locations where the detection significance exceeds the detection threshold), with most of the hotspots having relatively low detection significance. While many of the high detection significance hotspots correspond to known dwarf galaxies, the nature of the remaining hotspots are more ambiguous, with many likely being false positives caused by survey artifacts.

Due to the large number of hotspots, we initiated a Zooniverse citizen science project  ``DELVE Dwarf Galaxy Quest: Milky Way Neighbors'' to enlist volunteers to visually inspect the hotspots for high-quality galaxy candidates.\footnote{\url{https://www.zooniverse.org/projects/ywyh/delve-dwarf-galaxy-quest-milky-way-neighbors }\label{zooniverse_link}}  Zooniverse is the leading online platform for citizen science and currently has over 1.5 million ``citizen scientists" helping to analyze  more than 400 projects \citep{Fortson:2012, Zevin:2017}. The Zooniverse volunteers  are asked to examine diagnostic plots showing the smoothed stellar density map and color–magnitude diagram of regions around the  hotspots, similar to those shown in Figure~\ref{fig:diagnostic_plot}.\footnote{The volunteers are not given the Legacy Surveys Sky Viewer
images shown in the bottom row of Figure~\ref{fig:diagnostic_plot}.} They are then asked to classify each as either a promising candidate for follow-up or a likely false positive. Comprehensive results from this Zooniverse campaign, which will classify over 40,000 hotspots, will be presented in a forthcoming paper. 

Several promising candidates, however, were identified even before the launch of the Zooniverse campaign, during the preparation for the beta test. For this test, we generated diagnostic plots for a subset of $\sim$4,000 hotspots that passed the minimum detection threshold of ${\rm SIG_{gr}} > 3.0$ in \texttt{simple} and $\sqrt{\rm TS_{gr}} \gtrsim 3.17$ in \texttt{ugali}\footnote{Corresponding to a \texttt{ugali} threshold of ${\rm TS_{gr}} >10$},  with the spatial coincidence of the two detections within $0\fdg2$. 

Here, we present three of the most prominent satellite candidates identified: Carina~IV \CHECK{(${\rm SIG_{gr}} = 7.5$; $\sqrt{\rm TS_{gr}} = 8.8$)}, Phoenix~III \CHECK{(${\rm SIG_{gr}} = 4.9$; $\sqrt{\rm TS_{gr}} = 6.1$)}, and DELVE~7 \CHECK{(${\rm SIG_{gr}} = 4.9$; $\sqrt{\rm TS_{gr}} = 7.5$)}.  Carina IV and DELVE~7 \cytwo{were} first identified by DELVE members, \cytwo{based on the presence of clear Milky Way satellite signatures in their diagnostic plots (Fig.~\ref{fig:diagnostic_plot}), including a prominent overdensity in the stellar density map and a well-defined color--magnitude sequence consistent with a metal-poor stellar population. In contrast, Phoenix~III was discovered by a volunteer beta tester when going through the same diagnostic plots during the same phase.}\footnote{Phoenix~III was first identified by Zooniverse beta tester and citizen scientist Ernest Jude P.\ Tiu.} These candidates generally exhibit much higher \texttt{simple} and \texttt{ugali} detection significances compared to the remaining $\sim$4,000 hotspots detected by both algorithms, placing them within the top 6th percentile in \texttt{ugali} detection significance. 

We note that none of the three candidates exceeds the conservative detection thresholds adopted for inclusion in the systematic Milky Way satellite census of \citet{DELVECensus1}.\footnote{The detection threshold is ${\rm SIG_{gr}} \geq 5.5$, $\sqrt{\rm TS_{gr}} \geq 5.0$, and $\sqrt{\rm TS_{gi}} \geq 5.0$ within the DES footprint, and ${\rm SIG_{gr}} \geq 6.5$, $\sqrt{\rm TS_{gr}} \geq 8.0$, and $\sqrt{\rm TS_{gi}}\geq 7.5$ within the DELVE footprint, where  $\sqrt{\rm TS_{gi}}$ is the \texttt{ugali} detection significance in $g$,$i$ bands instead of $g$,$r$ bands. While Carina~IV satisfies the census thresholds for ${\rm SIG_{gr}}$ and $\sqrt{\rm TS_{gr}}$, it falls short of the $\sqrt{\rm TS_{gi}}$ criterion with $\sqrt{\rm TS_{gi}} = 6.4$. } \cytwo{In contrast to \citet{DELVECensus1}, which prioritizes high dwarf galaxy sample purity through the use of additional $i$ band information, the search and the main analysis presented in this work rely solely on $g$- and $r$-band data.}

\section{Characterization of the New Systems}
\label{sec:measurement}
In this section, we present measurements of the structural properties and stellar populations of the three most promising candidates from our search. These measurements are derived primarily with the maximum-likelihood framework of \texttt{ugali}, using the same DECam-based DELVE DR3 data employed in the search (Section~\ref{sec:ugali_measure}). In addition, Section~\ref{sec:imacs} details follow-up Magellan/IMACS imaging obtained for the faintest system, DELVE~7, which provides tighter constraints on its structural parameters and confirms its nature as a metal-poor stellar system.

\subsection{Properties from DECam Photometry}
\label{sec:ugali_measure}
We use the maximum-likelihood-based \texttt{ugali} codebase \citep{Bechtol:2015, MWCensus1} to measure the morphological and stellar population properties of the three candidates in the DELVE DR3 data. In the satellite search described in Section~\ref{sec:search}, we adopted a radially symmetric Plummer profile to reduce computation time, drawing parameters such as size and distance from a predefined grid and fixing a single composite isochrone. Here, we run \texttt{ugali} with a more flexible model that allows both the morphological and stellar population properties to vary freely.

We model the stellar density profile of each system with an elliptical Plummer profile, with the free parameters described by the centroid coordinates ($\alpha_{J2000},\delta_{J2000}$), angular semi-major axis length, $a_h$, ellipticity, $\epsilon$, and the position angle (P.A.) of the major axis (defined East of North). We then model the magnitudes and colors of the putative member stars with a \texttt{PARSEC} isochrone model \citep{Bressan:2012, Chen:2014, Tang:2014, Chen:2015} with the distance modulus, $(m-M)_0$, age, $\tau$, and metallicity, $Z_{\rm phot}$, of the system as parameters. We also fit the stellar richness, $\lambda$, which normalizes the total number of stars in the system \citep{Bechtol:2015, MWCensus1}.  

Since the three systems exhibit very different properties and have varying numbers and types of observable stars, we treat the isochrone parameters differently for each system. For the brightest system, Carina~IV, which has the largest number of observable stars, we allowed all the isochrone parameters ($(m-M)_0$, $\tau$, $Z_{\rm phot}$) to vary, finding a distance modulus of $(m-M)_0 = 20.1_{-0.2}^{+0.1}$,   age of $\tau = 12.5_{-1.0}^{+0.3}$ and a metallicity of $Z_{\rm phot} = 0.00013_{-0.00003}^{+0.00004}$. In the case of Phoenix~III, freeing all parameters led to a bimodal posterior, with one peak at $Z_{\rm phot} = 0.0001$, $\tau = 12$\,Gyr, and another more prominent peak at $Z_{\rm phot} = 0.0006$, $\tau = 7$\,Gyr. \cytwo{In the main analysis, we fixed the metallicity of the system to $Z_{\rm phot} = 0.0001$, and we obtained a distance modulus of $(m-M)_0 = 20.3_{-0.1}^{+0.2}$, age of $\tau = 11.8_{-0.4}^{+0.5}$\,Gyr. We discuss the physical motivation for this choice in Section~\ref{sec:phoenixIII} and examine the alternative younger solution in more detail in Appendix~\ref{appendix:phx3sc}.}  For DELVE~7, we adopted the best-fit isochrone from deeper Magellan/IMACS photometry, with $\tau = 10$\,Gyr, $Z = 0.0006$, and $(m-M)_0 = 18.1 \pm 0.3$ (see Section \ref{sec:imacs}).

We run the Markov Chain Monte Carlo sampler \texttt{emcee} \citep{Foreman_Mackey:2013}  for 5,000 steps with 20 walkers to simultaneously sample the structure and isochrone parameters in addition to the stellar richness. Table~\ref{table:params} shows the values and uncertainties of the stellar density profile and isochrone parameters obtained from \texttt{ugali}. The estimates of parameters are obtained from the median of the marginalized posteriors, while the uncertainties are obtained using the 16th and 84th percentiles.

Table \ref{table:params} also includes properties derived from the fitted parameters. For example, from the angular semi-major axis length, $a_h$, we can obtain the azimuthally-averaged angular half-light radii ($r_h$ = $a_h \sqrt{1-\epsilon}$). When combined with the distance modulus, we can derive the physical semi-major axes (in parsec), $a_{1/2}$, and azimuthally-averaged physical half-light radii (in parsec), $r_{1/2}$. Following the prescription of \citet{Martin2008}, we estimate the absolute $V$-band magnitude ($M_V$) and use them to derive the $V$-band luminosities ($L_V$). We derive stellar masses ($M_\star$) \CHECK{by assuming a stellar-mass-to-light ratio of $M_\star/L_V=2$} \citep{Simon:2019}.

For each star, the \texttt{ugali} pipeline assigns a probability that the star is a member of the stellar system based on its spatial position, photometric properties, and local imaging depth assuming a given model that includes a putative dwarf galaxy and the local stellar field population \citep{Bechtol:2015, MWCensus1}.
We plot the spatial distribution of stars in a small region around the three candidates, with stars colored by their \texttt{ugali} membership probability in the top panels of Figure~\ref{fig:ugali}. In the bottom panels of the same figure,  we show color--magnitude diagrams for each system with stars colored by their \texttt{ugali} membership probability and the best-fit \texttt{PARSEC} isochrone model \citep{Bressan:2012, Chen:2014, Tang:2014, Chen:2015}.

\begin{deluxetable*}{llcccc}
\tabletypesize{\scriptsize}
\tablewidth{0pt} 
\tablecaption{Measured and derived parameters of Carina IV, Phoenix III and DELVE 7. Details of each parameter can be found in their corresponding sections. \label{table:params} }
\tablehead{
\colhead{Parameter} & \colhead{Description} & \colhead{Units} & \colhead{Carina IV} & \colhead{Phoenix III} & \colhead{DELVE 7}   } 
\startdata
$\alpha_{J2000}$& Right Ascension of Centroid & deg & $104.046_{-0.015}^{+0.012}$ & $26.711_{-0.009}^{+0.008}$ & $304.107_{-0.005}^{+0.004}$  \\ 
$\delta_{J2000}$& Declination of Centroid & deg & $-63.319_{-0.005}^{+0.006}$   & $-41.428_{-0.003}^{+0.004}$  & $-50.33_{-0.002}^{+0.002}$  \\ 
$-$ & IAU Name & $-$ & J0656-6319 & J0146-4125 & J2016-5019  \\ 
$-$ & Survey Region$^a$ & $-$ & DELVE & DES & DES  \\ 
$a_h$ &  Angular Semi-Major Axis Length & arcmin  & $1.34_{-0.38}^{+0.56}$  & $0.85_{-0.37}^{+0.61}$  & $0.32_{-0.16}^{+0.58}$  \\
$a_{1/2}$&  Physical Semi-Major Axis Length &  pc  & $40_{-10}^{+20}$ & $28_{-12}^{+20}$ & $3_{-2}^{+7}$ \\ 
$r_h$&  Azimuthally-Averaged Angular Half-Light Radius&  arcmin & $1.19_{-0.34}^{+0.49}$ & $0.59_{-0.26}^{+0.42}$ & $0.2_{-0.1}^{+0.35}$  \\
$r_{1/2}$&  Azimuthally-Averaged Physical Half-Light Radius  & pc & $ 40_{-10}^{+10}$ & $ 19_{-8}^{+14}$ & $ 2_{-1}^{+4}$  \\ 
$\epsilon$&  Ellipticity  &  - & $0.21_{-0.21}^{+0.27}$ & $0.52_{-0.45}^{+0.18}$ & $0.63_{-0.53}^{+0.17}$   \\
P.A.&  Position Angle of Major Axis (East of North) & deg & $117_{-68}^{+63}$ & $89_{-40}^{+37}$ & $101_{-55}^{+43}$ \\ 
$(m-M)_0$&  Distance Modulus$^b$ & mag & $20.1_{-0.2}^{+0.1}$ & $20.3_{-0.1}^{+0.2}$ & $18.1_{-0.3}^{+0.3}$ \\
$D_\odot$&  Heliocentric Distance & kpc & $102_{-9}^{+5}$  & $115_{-7}^{+10}$  & $42_{-5}^{+6}$  \\ 
$D_{\rm GC}$&  Galactocentric Distance & kpc & $102_{-8}^{+2}$  & $115_{-5}^{+8}$ & $35_{-5}^{+6}$  \\ 
$M_V$&  Absolute (Integrated) $V$-band Magnitude  &  mag   & $-2.8_{-0.6}^{+0.4}$ & $-1.2_{-1.1}^{+0.7}$ & $1.2_{-2.1}^{+0.7}$ \\ 
$L_V$&  $V$-band  Luminosity &  L$_\odot$  & $1130_{-420}^{+590}$ & $260_{-140}^{+330}$  & $30_{-20}^{+60}$  \\
$M_*$& Stellar Mass  & M$_\odot$  & $2250_{-830}^{+1180}$  & $520_{-290}^{+660}$  & $60_{-40}^{+120}$ \\
$\tau$ & Best-Fit Age & Gyrs & 12.5  & 11.8 & 10.0  \\
$Z_{\rm phot}$ & Best-Fit Photometric Metallicity & dex & 0.00013  & 0.00010 & 0.00061  \\
$E(B-V)$ & Galactic Extinction & --- & 0.090  & 0.015 & 0.052  \\[+0.5em]
\hline
\enddata
\tablenotetext{a}{Here we refer to the generally deeper and more homogeneous DES portion of DELVE DR3 as the DES footprint, and the non-DES portion as the DELVE footprint.}
\tablenotetext{b}{Following \citet{Drlica-Wagner:2015}, we added in quadrature a 0.1 mag systematic uncertainty  to the distance modulus measurement to account for the uncertainties in the isochrone modeling.  }
\end{deluxetable*}

\subsection{Properties of DELVE~7 from Magellan/IMACS Photometry}
\label{sec:imacs}

\begin{figure*}[t!]
    \centering
    \includegraphics[width=0.45\textwidth]{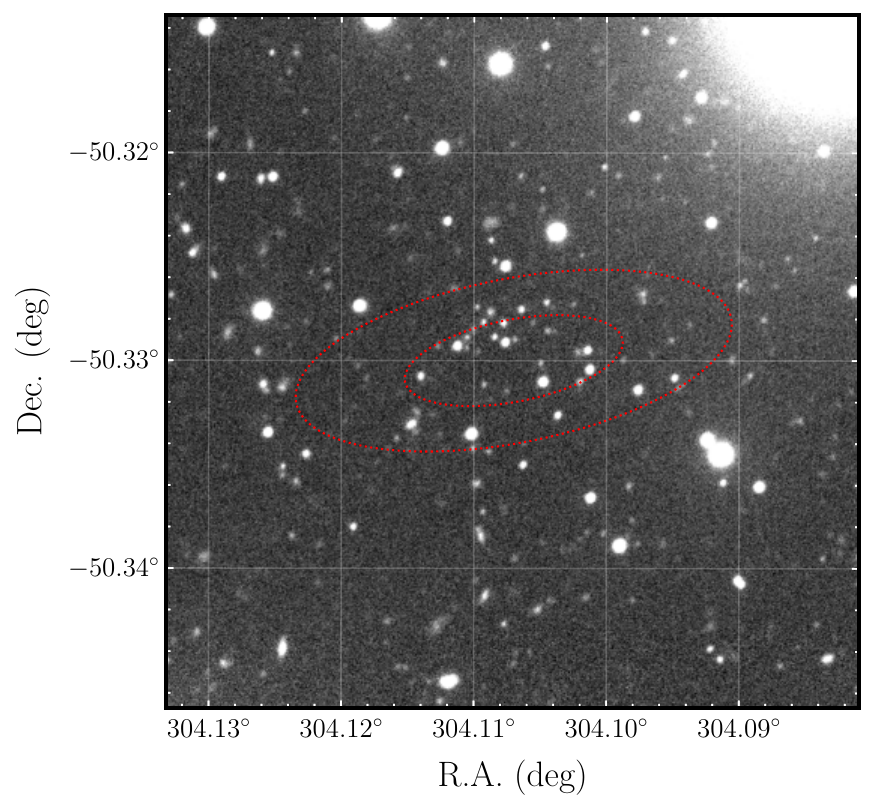}
    \includegraphics[width=0.54\textwidth]{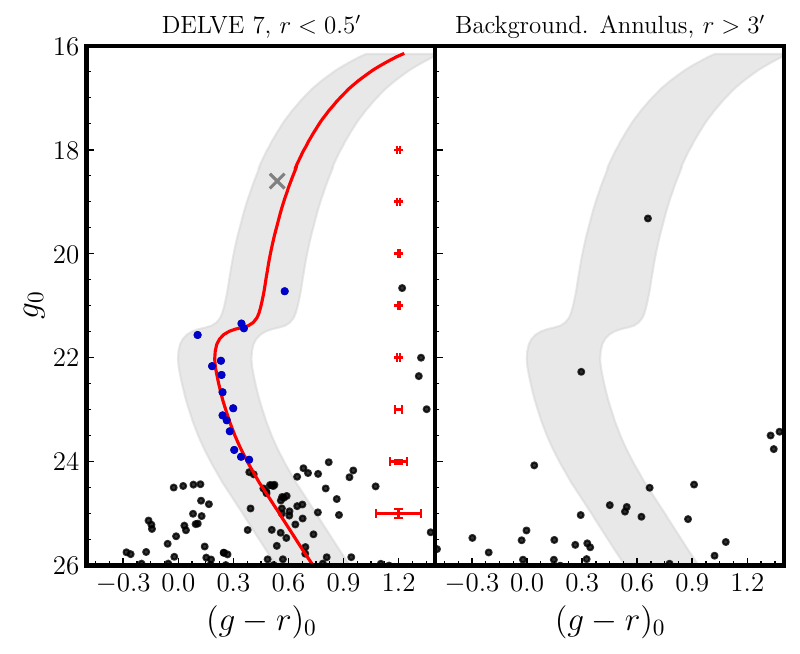}
    \caption{\label{fig:imacs} Follow-up imaging of DELVE~7 performed with Magellan/IMACS. (Left) \CHECK{$2\arcmin \times 2\arcmin$} $g$-band cutout image centered on DELVE~7, which can be seen as a clear overdensity of faint point-like sources. \CHECK{We overlay red ellipses demarcating $r_h$ and $2r_h$}. (Right) Color--magnitude diagrams for point-like sources in the vicinity of DELVE~7. The left panel shows a circular region with $r <  0\farcm5$ while the right panel shows a equal-area background annulus with inner radius $r=3\arcmin$. Stars used for the isochrone fit are shown in blue; the best-fit isochrone (shown in red) was found to be $\tau =10$~Gyr, $Z=0.00061$, $(m-M)_0 = 18.1$. \cytwo{The median Magellan photometric uncertainties for stars around DELVE 7, as a function of $g$-band magnitude, are indicated by red error bars.} Although one bright candidate RGB member falls close to the isochrone, this star has a \textit{Gaia} parallax of $\varpi = 0.71 \pm 0.15$~mas that places it well in the foreground of DELVE~7; we mark it here with an `x' to emphasize its non-membership. The shaded regions show $\pm 0.2$ in color about the best-fit isochrone.}
\end{figure*}
Unlike the two brighter objects identified in this paper, the classification of the faint DELVE 7 system as a real Milky Way satellite is not certain from DELVE DR3 alone. To further assess its nature, we obtained deep $g$- and $r$-band imaging using the Inamori Magellan Areal Camera \& Spectrograph \citep[IMACS;][]{Dressler:2006} on the 6.5-m Magellan Baade Telescope on 2025 April 4 during excellent conditions. 
We used the $f/2$ camera, which delivers a $\sim 27\farcm4$ field of view and $0\farcs2~{\rm pixel}^{-1}$ scale. 
Observations included $5 \times 300$\,s images in the $g$-band and $5 \times 300$\,s in the $r$-band, with small dithers between exposures. 
The data were reduced in a manner similar to \citet{Chiti:2020} and  \citet{Sand:2022}, which included overscan subtraction and flat-fielding, followed by an astrometric correction using a combination of \texttt{ASTROMETRY.NET} \citep{Lang:2010} and \texttt{SCAMP} \citep{Bertin:2006}. 
Image remapping and coaddition was performed with \texttt{SWARP} \citep{Bertin:2010} using a weighted average of the input images. 
The final $g$- and $r$-band stacked images have PSF FWHM values of $0\farcs 8$ and $0\farcs 7$, respectively. 
We display the final, stacked $r$-band IMACS image in Figure~\ref{fig:imacs} (left). DELVE~7 is clearly visible as a compact collection of faint stars at the center of the image.

We performed PSF photometry on the stacked IMACS images using \texttt{DAOPHOT} and \texttt{ALLFRAME} \citep{Stetson:1987, Stetson:1994} following the procedure described in \citet{MutluPakdil:2018}.
The photometry was calibrated to point sources in the DELVE DR3 catalog, including a color term, and was corrected for Galactic extinction following the procedure described in Section~\ref{sec:data}. 
The typical extinction at the position of DELVE~7 from \citet{Schlegel1998} is $E(B - V) = 0.052$, and we present the dereddened $g_0$ and $r_0$ magnitudes.
In Figure~\ref{fig:imacs} (right), we show the color--magnitude diagram of stars associated with DELVE 7 within $r < 0.5\arcmin$ along with the color--magnitude diagram of stars in an equal-area background annulus. Figure~\ref{fig:imacs} shows a sparse main sequence turn-off at $g \sim 21.5$ followed by a well-populated main sequence down to $g \sim 24.5$. Neither of these features are seen in the background region. We therefore confirm DELVE~7 as a bona fide new stellar system.

To obtain a more precise estimate of the properties of DELVE~7 from the IMACS imaging, we simultaneously fit the age, metallicity, and distance modulus of DELVE~7  using $\chi^2$ minimization over a restricted grid of PARSEC isochrone models spanning $10 < \tau < 13.5$ Gyr, $0.0001 < Z < 0.001$,  $17 < (m-M)_0< 19$. For each set of parameters, we computed a 1D $\chi^2$ comparing the model colors to the observed sample of stars within $r < 0\farcm5$ from the system centroid ($\approx \cytwo{2.5} \ r_h$, as estimated from \cytwo{the} \texttt{ugali} fit to the DECam data) with $g_0 < 24$, $(g-r)_0 < 1$. These cuts yield a pure selection of DELVE~7 members, shown in blue in Figure 
\ref{fig:imacs}.  The best-fit isochrone was found to be a $\tau = 10$~ Gyr, $Z=0.00061$, $(m-M)_0 = 18.09$~model. Given the extreme sparsity of the observed MSTO, however, the age and metallicity are largely unconstrained within our grid range, and thus these estimates are best considered nuisance parameters for the inference of the distance modulus. Based on the width of the main sequence in $g_0$ about the best-fit model, we visually estimated an approximate $\pm 0.3$~mag distance modulus uncertainty, i.e.,  $(m-M)_0 = 18.1 \pm 0.3$. This translates to a physical distance of $D_\odot = 42^{+6}_{-5}$~kpc.

\section{Proper Motion and Metallicity Measurements of Carina IV}
\label{sec:followup}
In this section, we present additional analysis that 
increase the confidence in the classification of Carina~IV as an old, metal-poor, bound stellar systems rather than chance alignments of Milky Way stars. In Section \ref{sec:gaia}, we use {\it Gaia} proper motion measurements to show that stars identified as Carina~IV members exhibit consistent motions. While in Section~\ref{sec:metallicities}, we use photometry from DECam $gri$ and CaHK filters to demonstrate that the members of Carina~IV have low metallicities, consistent with old stellar populations.

\begin{figure}
    \centering
    \includegraphics[width=\linewidth]{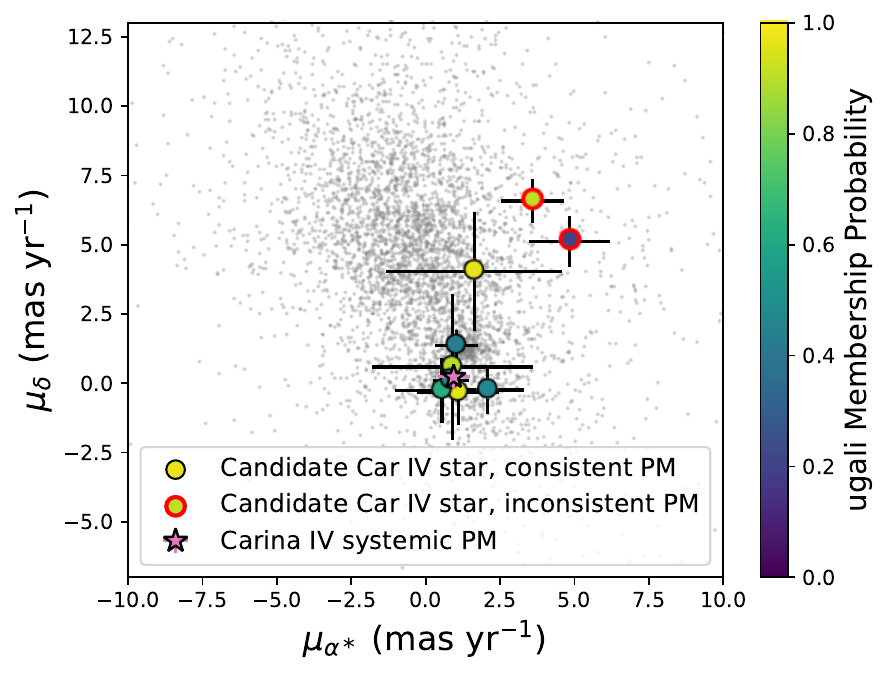}
    \caption{ Proper motion of candidate member stars in Carina IV using data from Gaia DR3.  The cluster of stars at $(\mu_\delta, \mu_{\alpha*}) \sim (0, 0)$\,mas/yr indicate the presence of a halo system. We outline in red the candidate stars whose proper motions deviate by more than 2$\sigma$ from the systemic Carina IV proper motion (shown as the pink star).  Stars within $0\fdg5$ of Carina IV  with a low \texttt{ugali} membership probability ($p_{\rm ugali} < 0.20$)  are shown in grey. \label{fig:carina4_pm}}
\end{figure}
\subsection{Proper Motion}
\label{sec:gaia}

We use  {\it Gaia} DR3 astrometry \citep{Gaia_Lindegren_2021A&A...649A...2L, Gaia_Vallenari2023A&A...674A...1G} to assist in the validation of our Milky Way satellite candidates. Of the three candidates, only Carina~IV has a clear detection in {\it Gaia} astrometry.  In Figure~\ref{fig:carina4_pm}, we compare the proper motion of stars with high  \texttt{ugali} membership ($p_{\rm ugali} > 0.2$) to the proper motion of all stars within \CHECK{$0\fdg5$}. There is a clear cluster of seven stars with low proper motion and zero parallax matching the red-giant branch of Carina~IV (See Table \ref{table:Carina4stars} in the Appendix for the  proper motion of individual stars).  These are also the seven closest stars \CHECK{with {\it Gaia} measurements} within 2$a_h$.  We note that the faintest star has $\texttt{ruwe}=1.3$ and $\texttt{astrometric\_excess\_noise\_sig}=2.3$, which is indicative of low-quality astrometry. 
The clear clustering of Carina~IV stars in {\it Gaia} confirms that it is a co-moving Milky Way satellite. Its low proper motion is also consistent with the stars being located at the distance of Carina~IV ($D_\odot = 102$ kpc).

Of the seven stars with {\it Gaia} proper motions that are consistent with Carina~IV, five of them are RGB stars and two are them Blue BHB stars. For the five RGB stars,  two of them have low metallicity (see next section) and are consistent with an ultra-faint dwarf galaxy origin. The other three stars have higher metallicity more consistent with a Milky Way or LMC origin. These three stars are the {\it Gaia} candidates with the largest separation from the centroid of Carina~IV and also have the lowest \textit{ugali} membership probability. 
A clear {\it Gaia} detection remains if these three RGB stars are removed as interlopers.
The systemic proper motion we measure from the four remaining confident {\it Gaia} members (two RGB stars and two BHB stars) is: $\mu_{\alpha *} = 0.94_{-0.53}^{+0.54}~{\rm mas~yr^{-1}}$, $\mu_{\delta} = 0.23_{-0.40}^{+0.41}~{\rm mas~yr^{-1}}$, with a correlation coefficient between $\mu_{\alpha *}$ and $\mu_{\delta}$ of $C_{\mu_{\alpha *} \times \mu_{\delta}}=0.19$.

For the other two candidates, the majority of the stars that possess high \texttt{ugali} membership probabilities ($p_{\rm ugali} > 0.2$) are fainter than the {\it Gaia} detection limit of $G \sim 21$, preventing us from obtaining {\it Gaia} proper motion measurements \cytwo{(see Figure~\ref{fig:ugali} and Appendix~\ref{appendix:carina4_stars} for more details)}. However, future datasets, such as Vera C.\ Rubin Observatory’s Legacy Survey of Space and Time (LSST; \citealt{Ivezic:2019}), may provide proper motion measurements for these fainter stars.

\subsection{  Photometric Metallicities}
\label{sec:metallicities}

Carina~IV was also imaged with the metallicity-sensitive, narrowband CaHK filter (N395) on DECam as part of the ongoing The Mapping the Ancient Galaxy in CaHK (MAGIC) survey (Chiti et al.\ in prep.). 
The N395 filter covers the Ca\,{\sc ii} H \& K lines and is similar to other metallicity sensitive filters \citep[e.g., as used by the Pristine survey;][]{Starkenburg2017MNRAS.471.2587S}. 
To derive metallicities, the survey follows the same approach as in \citet{Chiti:2020, Chiti:2021}, which involves forward-modelling  fluxes using a grid of synthetic spectra generated with the Turbospectrum code \citep{Alvarez:1998, Plez:2012} under the assumption of Local Thermodynamic Equilibrium (TSLTE). Further details on the TSLTE grid can be found in \citet{Barbosa2025arXiv250403593B}. Analysis of the Sculptor dwarf spheroidal galaxy with MAGIC has determined that the metallicity precision achievable with the N395 filter in combination with DELVE broadband imaging can be as low as $\sigma_{\rm [Fe/H]}\sim 0.16$ \citep{Barbosa2025arXiv250403593B}, while the first high-resolution spectroscopic follow-up of six MAGIC candidates found a median $\Delta \rm{[Fe/H]}=-0.06$, confirming the high precision of the photometric estimates \citep{Placco:2025}.

\begin{figure}
    \centering
    \includegraphics[width=\linewidth, clip, trim={0.5cm 0 0.5cm 0cm}]{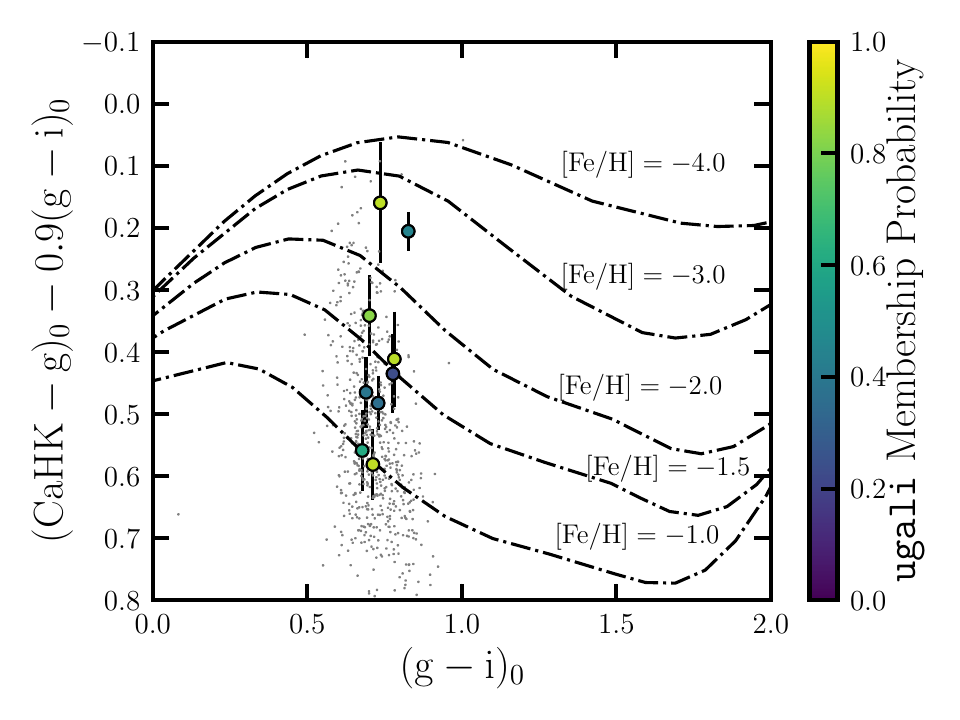}
    \caption{\label{fig:cahk_carina4} Color--color diagram for possible RGB members of Carina IV (colored by their \texttt{ugali} membership probability) and the Milky Way foreground  (gray points) now including narrow-band CaHK observations from the MAGIC survey. The dot-dashed lines show iso-metallicities lines
     for synthetic RGB stars with  ${\rm [Fe/H]} = {-1.0, -1.5, -2.0, -3.0, -4.0}$, assuming a surface gravity of $\log g = 2$ and effective temperatures ranging from 3500 to 7500 K, generated from the TSLTE grid. The detection of four low-metallicity stars  (${\rm [Fe/H]} < -1.5$) confirms Carina IV as an old, metal-poor system.}
\end{figure}

Carina~IV was serendipitously observed by MAGIC with a single $1 \times 720$\,s image in the N395 filter, and an additional $2 \times 720$\,s dithers were obtained after Carina~IV was identified as a dwarf galaxy candidate. For the N395 imaging, we are able to estimate metallicities close to the \textit{Gaia} detection limit, achieving uncertainties of $\sigma_{\rm [Fe/H]} \sim 0.5$ at $g \sim 21$. To verify the identity of Carina~IV, we examine the MAGIC metallicities for stars with high \texttt{ugali} membership probability ($p_{\rm ugali} > 0.2$). As shown in Figure~\ref{fig:cahk_carina4}, we find four stars with high \texttt{ugali} membership and low MAGIC metallicity (${\rm [Fe/H]} < -1.5$) similar to other ultra-faint dwarf galaxies \citep[e.g.,][]{Simon:2019, Fu:2023}. \cytwo{The remaining five stars with MAGIC metallicity measurements are possibly contaminants from either the LMC halo or the Milky Way halo.}  The mean metallicity determined by MAGIC for these four stars is $\overline{\rm [Fe/H]}=-2.0\pm0.2$, which is on the high end of the ultra-faint dwarf galaxy population \citep{Fu:2023}. 
However, the detection of low-metallicity stars confirms Carina~IV as an old, metal-poor system and motivates spectroscopic follow up. The individual metallicity measurements for the high-\texttt{ugali}-membership stars are provided in Table \ref{table:Carina4stars} in Appendix \ref{appendix:carina4_stars}. 

\section{Discussion }
\label{sec:discussion}

\begin{figure*}
    \centering
    \includegraphics[width=\linewidth]{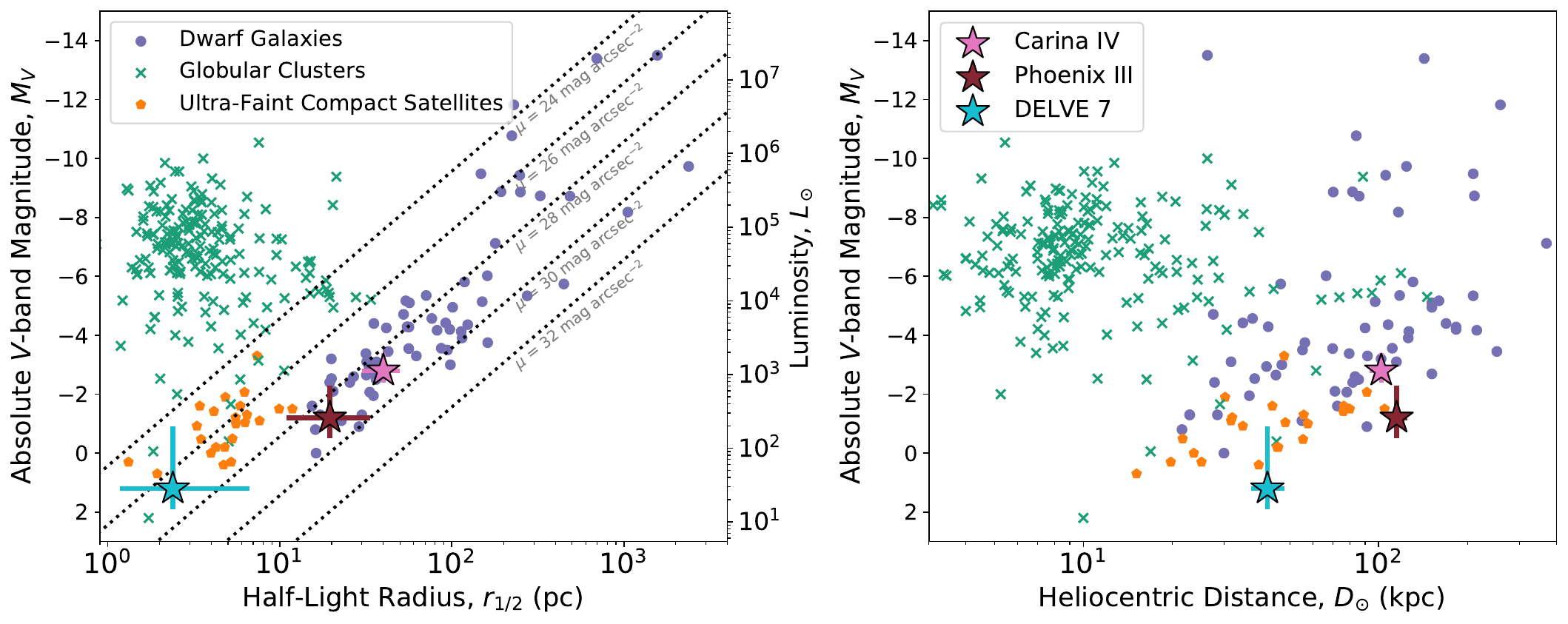}
    \caption{\label{fig:size_lum} (Left) Absolute $V$-band magnitude ($M_V$) versus azimuthally averaged physical half-light radius ($r_{1/2}$) for the three new Milky Way satellites and the population of known Local Group globular clusters and dwarf galaxies. (Right) The same diagram, but showing $M_V$ versus heliocentric distance ($D_{\odot}$). Spectroscopically confirmed dwarf galaxies and likely galaxy candidates are shown in purple circles, while Milky Way globular clusters are shown with green `$\times$' symbols. Ambiguous ultra-faint compact satellites are shown with orange pentagon symbols. The three systems discovered in this work are shown as colored stars with bars representing 68\% posterior uncertainty intervals. Data on the Milky Way satellite population were compiled by \citet{Pace:2024}, and we adopt the same classifications as \citet{Tan:2025b}.}
\end{figure*}

In this section, we highlight notable features of the three newly discovered Milky Way satellite candidates. As outlined in the introduction, there is ongoing debate about the classification of the faintest Milky Way satellites, collectively referred to as ultra-faint compact satellites, particularly whether they are dark-matter dominated dwarf galaxies or baryon-dominated star clusters. In light of this debate, we present our classification of the three systems and explain the reasoning behind our interpretation. In Figure~\ref{fig:size_lum}  we compare the sizes and luminosities of our three new satellites with both Milky Way satellite dwarf galaxy and star cluster population.

\subsection{Carina~IV}
\label{sec:carinaIV}
Carina~IV is the most luminous of the three candidates presented here and is the best-characterized based on the available data. The diagnostic plots in Figure~\ref{fig:diagnostic_plot} show a well-populated RGB, consistent with an old, metal-poor stellar population. In addition, the stars with high \texttt{ugali} membership probabilities exhibit proper motions clustered around $(\mu_{\alpha*}, \mu_\delta) \sim (0, 0)$\,mas\,yr$^{-1}$, which is consistent with a distant, gravitationally bound Milky Way satellite. The low metallicity revealed through CaHK imaging  supports the interpretation that Carina~IV  as an old, metal poor system.

Overall, Carina~IV's properties are consistent with confirmed satellite galaxies toward the faint end of the luminosity function. Specifically, Carina~IV's extended physical size ($r_{1/2} = 40$\,pc) is larger than almost all known star clusters.   One notable feature of Carina~IV is its proximity in projection to the Magellanic Clouds, lying at an angular distance of $\sim 11\fdg6$ from the LMC, raising the possibility of an association, with 7 other systems already confirmed to be LMC satellites \citep{Kallivayalil:2018, Erkal:2020,  Patel:2020, 2022MNRAS.511.2610C, 2024MNRAS.527..437V,  Pace:2025}.

\cytwo{We note in Fig. \ref{fig:diagnostic_plot} and \ref{fig:ugali} an excess of sources just above the Carina IV MSTO at $(g-r)_0 \sim 0.2$ and $g \sim 23$. Given Carina IV’s close proximity to the LMC, and the fact that these stars align well with the LMC main sequence, we interpret these sources as likely LMC halo main-sequence stars. This interpretation is further supported by the observation that when we apply an isochrone cut based on the LMC isochrone, we observe a background gradient with increasing density toward the direction of the LMC.}

To test the potential association of Carina~IV with the LMC, we carry out orbit modeling in a combined Milky Way + LMC potential following the methods of \citet{Erkal:2020}.  
We use the \citet{McMillan2017MNRAS.465...76M} potential for the Milky Way and for the LMC we assume a Hernquist profile with a total mass of $1.38 \pm 0.255 \times 10^{11}\,{\rm M_\odot}$ \citep{Erkal2019MNRAS.487.2685E} and a scale radius chosen to match the enclosed mass of $1.7\times10^{10}\,{\rm M_\odot}$ at 8.7\,kpc from \citet{vanderMarel2014ApJ...781..121V}.
As the radial velocity of Carina~IV is unknown, we vary the velocity between $-200$ to 800\,km\,s$^{-1}$ in steps of 5\,km\,s$^{-1}$.  At each velocity we draw 100 samples from the Carina~IV distance and proper motion errors, the LMC phase space and total mass, and the posterior for the Milky Way potential. 

We find a low probability of 1.8\% for Carina~IV to be associated with the LMC based on our orbit modeling. To evaluate this probability, we count the number of realizations in which Carina~IV was energetically bound to the LMC at the end of orbit rewinding following \citet{Erkal:2020}.
Only the velocity range of $\sim200-450~{\rm km~s^{-1}}$ has a multitude of bound samples.  In this velocity range, the association probability varies between 2\% and 9\%, and the total probability is 7\%.
In the orbital models, we find an average relative distance of $49_{-9}^{+21}\,{\rm kpc}$ and relative velocity of $ 373_{-151}^{+153}$\,km\,s$^{-1}$ with respect to the LMC at Carina IV's orbital pericenter with respect to the LMC, while the LMC's escape velocity at $49~{\rm kpc}$ is $\sim135$\,km\,s$^{-1}$.
We note that the large tangential velocity errors ($\approx 200$\,km\,s$^{-1}$) make it difficult to accurately determine the orbit and any association with the LMC; however, these errors are expected to decrease by a factor of $\sim$2.8 in the upcoming {\it Gaia} DR4. 
Overall, Carina~IV is more likely to be a Milky Way satellite behind the LMC system than to be part of the LMC system.

\subsection{Phoenix III}
\label{sec:phoenixIII}

\par Due to its low luminosity, large heliocentric distance, and the absence of follow-up observations, Phoenix~III is the least well characterized of the three newly discovered systems presented in this paper. In fact, at a heliocentric distance of \CHECK{$D_\odot = 115^{+10}_{-7}$\,kpc} and an absolute $V$-band magnitude of \CHECK{$M_V = -1.2^{+0.7}_{-1.1}$}, it is the faintest system discovered at a distance of \CHECK{${>}100$\,kpc} to date \citep{Pace:2024}.
The dwarf galaxy candidates Sextans~II ($D_\odot = 126$\,kpc; $M_V = -3.9$) and Virgo~III ($D_\odot = 151$\,kpc; $M_V = -2.7$) were discovered at slightly farther distances in data from KiDS and HSC-SSP, but are both considerably brighter \citep{Homma:2024,Gatto:2024}.
The comparably faint Kim 2 ($D_\odot = 105$\,kpc; $M_V = -1.5$) is slightly closer and was discovered in early DECam data and confirmed by deep Gemini/GMOS imaging \citep{Kim:2015}. Also comparable is the fainter, but slightly closer, Virgo~I ($D_\odot = 89$\,kpc; $M_V = -0.7$), which was first identified in the deeper HSC-SSP survey (\citealt{Homma:2016}; Crnojević et al.\ in prep.).

The depth and uniformity of DES Y6 make it possible to measure half-a-dozen stars in the subgiant branch of Phoenix~III and produce a relatively clean identification of a spatial overdensity.
Visual inspection of color images from the Legacy Surveys Sky Viewer shows a collection of faint blue stars (Figure~\ref{fig:diagnostic_plot}); however, the larger angular size of Phoenix~III (\CHECK{$a_h = 0\farcm85$}) makes it difficult to identify visually.

\cytwo{As discussed in Section~\ref{sec:ugali_measure}, the morphological fits for Phoenix~III yield two distinct posterior peaks. One corresponds to an old stellar population with an age of $\sim$12~Gyr, while the other favors an intermediate-age population of $\sim$7~Gyr. The two solutions have similar structural parameters, including total magnitude, physical size, and heliocentric distance. We favor the old, metal-poor $\sim$12~Gyr solution, consistent with other old Milky Way ultra-faint satellites \citep{Durbin:2025}. This preference is motivated by the fact that the metal-poor isochrone allows for the inclusion of the candidate blue horizontal branch (BHB) star, \textit{Gaia} DR3 4959457079527646080 (see also Figure \ref{fig:alt_phx3}). The star has colors $g_0 - r_0 = -0.21 \pm 0.01$ and $i_0 - z_0 = -0.05 \pm 0.02$, which are consistent with a BHB classification from \citet{Belokurov:2016}, albeit with large photometric uncertainties due to its faintness. Moreover, using the method of \citet{Belokurov:2016} to derive an estimate of $M_g$, we find that the BHB candidate has a distance modulus of $(m-M)_0 = 20.3$, which is consistent with that of Phoenix~III.  It is unlikely that a random Milky Way field BHB star is contaminating Phoenix~III, as field BHBs are rare at such large Galactocentric distances \citep[e.g.,][]{Fukushima:2018,Thomas:2018,Yu:2024}. However, since only a single BHB candidate is identified, we cannot rule out the possibility that it is instead a blue straggler or another type of contaminant. Therefore, for completeness, we also present the isochrone fits corresponding to the intermediate-age solution in Appendix~\ref{appendix:phx3sc}.} 

The median physical size (\CHECK{$r_{1/2} = 19$\,pc}), absolute magnitude (\CHECK{$M_V = -1.2$}), and heliocentric distance (\CHECK{$D_\odot = 115$\,kpc}) place Phoenix~III in a region of parameter space that is occupied by old, metal-poor dwarf galaxies and star clusters, though it is more heavily populated by the former (Figure~\ref{fig:size_lum}). \cytwo{This motivated our classification of the system as a likely dwarf galaxy.} However,  Kim 2 ($D_\odot = 105$\,kpc, $M_V = -1.5$, $r_{1/2} = 12$\,pc) possesses similar properties and is suspected to be a star cluster \citep{Kim:2015}. Moreover, given the large uncertainties in the size measurement of Phoenix~III, the possibility that it is a star cluster cannot be ruled out.

Despite residing in the DES footprint, Phoenix~III was overlooked by previous searches of the DES data \citep{Bechtol:2015, Koposov15a, Drlica-Wagner:2015, Kim:2015b, MWCensus1, Mcnanna:2024}. This is due to the importance of the faint sub-giant branch stars at $g \sim 24$, which only became accessible with the most recent DES data release. In searches of the DES Y3 data, a hotspot was identified at the location of Phoenix~III with a significance of $\sqrt{\rm TS} = 4.9$ and ${\rm SIG} = 5.5$, which did not exceed the threshold of that search ($\sqrt{\rm TS} > 6$, SIG $>$ 6).
It is likely that deeper imaging over this region of the sky by Rubin LSST will yield more systems similar to Phoenix~III \citep{Tsiane:2025}.

\subsection{DELVE~7}
\label{sec:delve7}
DELVE~7, located in the constellation Telescopium, is the faintest system \CHECK{($M_V = +1.2^{+0.7}_{-2.1}$)} among the three candidates identified in this analysis. Though sparse, this candidate was distinguished from the other less prominent candidates through visual inspection of the Legacy Surveys Sky Viewer DR10 coadded images, which reveal a clear clustering of blue stars (Figure~\ref{fig:diagnostic_plot}). As with Phoenix~III, a hotspot was identified at the location of DELVE~7 in searches of the DES Y3 data with a significance of \CHECK{$\sqrt{\rm TS} = 6.2$ and ${\rm SIG} = 5.3$}, which was lower than the detection threshold applied to that earlier search ($\sqrt{\rm TS} > 6$, SIG $>$ 6; \citealt{MWCensus1}). 

\par While the identity of DELVE~7 as a {\it bona fide} stellar system was uncertain based on the DECam imaging alone, the follow-up Magellan/IMACS imaging reveals a sparse but unambiguous main sequence extending to $g_0\sim24.5$. Our \texttt{ugali} fit of the DECam data suggests that DELVE~7 is the faintest Milky Way satellite discovered beyond the inner Milky Way stellar halo ($D_{GC}>20$ kpc; \citealt{Carollo:2007}). Its luminosity is bracketed on the faint end by Ursa Major~III/UNIONS~1 and on the bright end by Kim~3 ($M_V = +0.7$) and DELVE~5 ($M_V = +0.4$). Like  these other faint satellites, DELVE~7 is physically compact \CHECK{($r_{1/2} = 2$\,pc)} and features at most one RGB star (and more likely none) owing to the stochastic sampling of the IMF in low-mass systems. 

\par DELVE~7 is broadly representative of the population of ultra-faint compact satellites discovered in recent years \citep[e.g.,][]{2012ApJ...753L..15M,Kim:2015,2015ApJ...799...73K,Mau:2019, Cerny:2023c,Cerny:2023b,Smith:2024}. While past convention has often been to refer to these systems as ultra-faint star clusters, recent investigations into the population of Milky Way satellites at $r_{1/2} < 15$\,pc have tentatively suggested that some of these systems may in fact be galaxy candidates \citep{Simon:2024, Cerny:2026arXiv260217652C}. Thus, in the absence of any spectroscopic information about the system's internal kinematics and metallicity distribution, either a dwarf galaxy or a star cluster classification remains possible for DELVE~7. 
However, because no galaxies have yet been confirmed in this size or luminosity regime, we conservatively assume that the system is more likely to be a star cluster and name it DELVE~7. 
\par If DELVE~7 is indeed a star cluster, it is likely that the system has lost a significant fraction of its birth mass. Even in the absence of Galactic tides, DELVE~7's dynamical relaxation timescale is just $\sim  70$\,Myr (calculated assuming its current stellar mass and radius of $M_* = 60\,{\rm M}_\odot$, $r_{1/2} = 2 \,{\rm pc}  $), which is far smaller than its observed age. Thus, it is expected that the system should be evaporating due to a combination of mass segregation and internal two-body interactions. Importantly, though, we find that it is not necessary to invoke the presence of unseen mass (e.g., dark matter or a substantial population of stellar remnants) to explain its long-term survival and the existence of a bound remnant of its mass. The evaporation timescale estimated from its present-day mass, $t_{\rm evap} \approx 140 t_{\rm relax} \sim 10$\,Gyr, should be regarded as a conservative lower limit, yet it already matches the age inferred from isochrone fitting. In other words, its evaporation timescale is sufficiently long  that DELVE~7 is not expected to have completely evaporated over its 10\,Gyr lifetime. This distinguishes the system from the single lower-mass halo satellite currently known, Ursa Major~III/UNIONS~1, where similar calculations suggested that the system should have totally evaporated ($\tau \gg t_{\rm evap}$) \citep{Errani:2024}.

Compared to the other candidates, DELVE~7 is relatively nearby, with a distance modulus of only $(m-M)_0 = 18.1 \pm 0.3$. This places the system’s horizontal branch above $g = 20$, a magnitude range where the \textit{Gaia} DR3 catalog of RR Lyrae stars is expected to be complete \citep{Clementini:2023}. Nevertheless, we found no RR Lyrae stars within $r < 2r_h$ of DELVE~7. This absence is not surprising given the lack of horizontal branch stars. Several other faint stellar systems are also known to lack RR Lyrae stars \citep{MartinezVazquez:2023}, and a fit to the specific frequency of RR Lyrae found in ultra-faint dwarf galaxies predicts $\ll 1$ RR Lyrae star in systems as faint as DELVE~7 \citep{Martinez-Vazquez:2019}.

\par The discovery of another relatively nearby \CHECK{($D_{\odot} < 50$\,kpc)} satellite at $M_V > 0$ emphasizes the incompleteness of current surveys for systems of this type. The recent DES Y6 survey sensitivity analysis performed in \citet{DELVECensus1} predicts that even with a low detection threshold of \CHECK{$\sqrt{\rm TS} > 5.5$} and {${\rm SIG} > 5.5$}, the 50\% detection completeness for systems like DELVE~7 in DES Y6 extends to only \CHECK{$\sim$80\,kpc}. The sensitivity to these systems should increase appreciably with the advent of Rubin LSST \citep{Tsiane:2025}. However, population-level predictions are still highly uncertain for systems of this type, given the unknown behavior of the luminosity function at $L < 10^3\,{\rm L_\odot}$ and the relatively unknown radial distribution of these faint satellites.

\section{Summary}
\label{sec:summary}

We present the discovery of three Milky Way satellite systems, Carina~IV, Phoenix~III, and DELVE~7, which were found in a search of DELVE DR3. Two of these systems, Carina~IV and Phoenix~III, have physical sizes and luminosities that place them in a region of parameters space that is consistent with the locus of ultra-faint dwarf galaxies, while the third system, DELVE~7, is more compact and consistent with the growing population of halo systems with small sizes and indeterminate classifications (Figure~\ref{fig:size_lum}). These objects were not included in the recent census of Milky Way satellites presented in \citet{DELVECensus1} due to the fact that their detection significances are lower than the thresholds set for that analysis.

Carina~IV, Phoenix~III, and DELVE~7 bring the total number of ultra-faint Milky Way satellites discovered in the DELVE data to 17 (\citealt{Mau:2020, Cerny:2021b, Cerny:2021, Cerny:2023c, Cerny:2023b, Cerny:2023,Cerny:2024, Tan:2025}). These three candidates were identified using the DELVE DR3 catalog, which provides deeper and more accurate photometric measurements than previous DELVE releases \citep{Drlica-Wagner:2021,Drlica-Wagner:2022}. Two of the candidates lie within the DES portion of the catalog, which was constructed from the full six years of DES Y6 observations and therefore provides generally deeper data than the rest of DELVE DR3. 

\cite{Manwadkar:2022} forecasted that the DELVE wide-area survey would find $64_{-13}^{+17}$ Milky Way satellites with $M_V < 0$ and $r_{1/2} > 10$\,pc at $\delta_{J2000} < 0^{\circ}$,\footnote{{DELVE subsequently expanded coverage to $\delta_{J2000} \lesssim 30^{\circ}$.}} 
while only \CHECK{36 systems} have been discovered in this region so far.  A more comprehensive search of DELVE DR3 is expected to yield additional satellite discoveries. One such effort to fully exploit DELVE data is the previously mentioned Zooniverse citizen-science project,$^{\ref{zooniverse_link}}$ in which volunteers visually inspect candidate systems. We expect to release the results of this project soon, once the public classifications are complete, the data have been analyzed, and the most promising candidates have been followed up.

Even with a complete search using data from the current surveys, we expect to be  largely incomplete for faint and distant systems such as those presented in this analysis. However, the upcoming Rubin LSST \citep{Ivezic:2019} is expected to discover hundreds of ultra-faint dwarf galaxies in the Local Volume \citep{Hargis:2014, Mutlu-Pakdil:2021, Manwadkar:2022, Tsiane:2025}. Space-based facilities, including the Roman Space Telescope \citep{Spergel:2015} and Euclid \citep{Euclid:2022}, are also anticipated to uncover additional ultra-faint dwarfs \citep{Nadler:2024}. This rapidly growing census of ultra-faint dwarfs will provide critical insight into the processes of galaxy formation at the smallest scales and serve as a powerful probe of dark matter physics.

\section*{Acknowledgment}
\cytwo{We thank the anonymous referee for the many useful comments that helped us improve this manuscript.}
We would also like to thank the Zooniverse team, beta testers, and volunteers for their support in setting up the Zooniverse beta test, and especially citizen scientist and beta tester Ernest Jude P.\ Tiu, who was the first to identify Phoenix~III during the beta test. For this work, we did not make use of the classifications made by other Zooniverse citizen scientists or beta testers. However, we plan to provide a more detailed description of those efforts in a forthcoming paper.

\defcitealias{McQuinn:2025hst..prop18066M}{HST GO Cycle 33 \#18066}
During the final stages of preparing this manuscript, we became aware that Phoenix~III had been independently identified by McQuinn, Mao, Buckley, Shih, Dolphin, Cohen, Tollerud, Hai, Leishman, and Brown as a candidate Local Group galaxy (\citetalias{McQuinn:2025hst..prop18066M}).

%
%

CYT was supported by the U.S.\ National Science Foundation (NSF) through grants AST-2108168 and AST-2307126.  WC gratefully acknowledges support from a Gruber Science Fellowship at Yale University.  This material is based upon work supported by the National Science Foundation Graduate Research Fellowship Program under Grant No.\ DGE2139841. DJS acknowledges support from NSF grant AST-2205863.

The DELVE project is partially supported by the NASA Fermi Guest Investigator Program Cycle 9 No.\ 91201 and by Fermilab LDRD project L2019-011. This material is based upon work supported by the National Science Foundation under Grant No.\ AST-2108168, AST-2108169, AST-2307126, and AST-2407526. This research award is partially funded by a generous gift of Charles Simonyi to the NSF Division of Astronomical Sciences.  The award is made in recognition of significant contributions to Rubin Observatory’s Legacy Survey of Space and Time.  Any opinions, findings, and conclusions or recommendations expressed in this material are those of the author(s) and do not necessarily reflect the views of the National Science Foundation.

Funding for the DES Projects has been provided by the U.S. Department of Energy, the U.S. National Science Foundation, the Ministry of Science and Education of Spain, 
the Science and Technology Facilities Council of the United Kingdom, the Higher Education Funding Council for England, the National Center for Supercomputing 
Applications at the University of Illinois at Urbana-Champaign, the Kavli Institute of Cosmological Physics at the University of Chicago, 
the Center for Cosmology and Astro-Particle Physics at the Ohio State University,
the Mitchell Institute for Fundamental Physics and Astronomy at Texas A\&M University, Financiadora de Estudos e Projetos, 
Funda{\c c}{\~a}o Carlos Chagas Filho de Amparo {\`a} Pesquisa do Estado do Rio de Janeiro, Conselho Nacional de Desenvolvimento Cient{\'i}fico e Tecnol{\'o}gico and 
the Minist{\'e}rio da Ci{\^e}ncia, Tecnologia e Inova{\c c}{\~a}o, the Deutsche Forschungsgemeinschaft and the Collaborating Institutions in the Dark Energy Survey. 

The Collaborating Institutions are Argonne National Laboratory, the University of California at Santa Cruz, the University of Cambridge, Centro de Investigaciones Energ{\'e}ticas, 
Medioambientales y Tecnol{\'o}gicas-Madrid, the University of Chicago, University College London, the DES-Brazil Consortium, the University of Edinburgh, 
the Eidgen{\"o}ssische Technische Hochschule (ETH) Z{\"u}rich, 
Fermi National Accelerator Laboratory, the University of Illinois at Urbana-Champaign, the Institut de Ci{\`e}ncies de l'Espai (IEEC/CSIC), 
the Institut de F{\'i}sica d'Altes Energies, Lawrence Berkeley National Laboratory, the Ludwig-Maximilians Universit{\"a}t M{\"u}nchen and the associated Excellence Cluster Universe, 
the University of Michigan, NSF NOIRLab, the University of Nottingham, The Ohio State University, the University of Pennsylvania, the University of Portsmouth, 
SLAC National Accelerator Laboratory, Stanford University, the University of Sussex, Texas A\&M University, and the OzDES Membership Consortium.

Based in part on observations at NSF Cerro Tololo Inter-American Observatory at NSF NOIRLab (NOIRLab Prop. ID 2012B-0001; PI: J. Frieman, NOIRLab
Prop. ID 2019A-0305; PI: Alex Drlica-Wagner, and NOIRLab Prop. ID 2023B-646244; PI: Anirudh Chiti), which is managed by the Association of Universities for Research in Astronomy (AURA) under a cooperative agreement with the National Science Foundation.

The DES data management system is supported by the National Science Foundation under Grant Numbers AST-1138766 and AST-1536171.
The DES participants from Spanish institutions are partially supported by MICINN under grants PID2021-123012, PID2021-128989 PID2022-141079, SEV-2016-0588, CEX2020-001058-M and CEX2020-001007-S, some of which include ERDF funds from the European Union. IFAE is partially funded by the CERCA program of the Generalitat de Catalunya.

We  acknowledge support from the Brazilian Instituto Nacional de Ci\^encia
e Tecnologia (INCT) do e-Universo (CNPq grant 465376/2014-2).

This document was prepared by the DES Collaboration using the resources of the Fermi National Accelerator Laboratory (Fermilab), a U.S. Department of Energy, Office of Science, Office of High Energy Physics HEP User Facility. Fermilab is managed by Fermi Forward Discovery Group, LLC, acting under Contract No.\ 89243024CSC000002.

This paper includes data gathered with the 6.5 meter Magellan Telescopes located at Las Campanas Observatory, Chile.

\section*{Author Contributions}
CYT performed the dwarf galaxy search, carried out the \texttt{ugali} morphology fits, produced most of the plots and tables, and led the writing of the paper. WC conducted the analysis of DELVE~7 based on the follow-up IMACS imaging, generated Figure~\ref{fig:imacs}, assisted with candidate identification, and contributed to the writing. ABP conducted the CaHK and proper motion analysis, contributed Figure~\ref{fig:cahk_carina4}, assisted with candidate identification, and contributed with writing.  JAS and KO helped identify dwarf galaxy candidates through preparation for the Zooniverse project, \cytwo{and contributed to  Figure~\ref{fig:diagnostic_plot}}. ADW provided direct supervision of the research and contributed with writing. JDS carried out the IMACS follow-up observations of DELVE 7, while DJS and BMP reduced the imaging data and produced the catalogs.
AMS performed the orbital modeling of Carina IV. 
DE, PSF, FS internally reviewed the paper. The authors from KRA to AKV contributed to producing and characterizing one or more of the following data products used in the paper: DES Y6 source catalog, DELVE DR3 source catalog, MAGIC catalog, and/or provided valuable comments that improved the paper’s clarity and quality. Builders: The remaining authors contributed to this work through the construction of DECam and other aspects of data collection; data processing and calibration; developing widely used methods, codes, and simulations; running pipelines and validation tests; and promoting the science analysis.



%

\vspace{5mm}
\facilities{Blanco/DECam, Magellan/IMACS, {\it Gaia}}


\software{
\texttt{astropy} \citep{2013A&A...558A..33A, 2018AJ....156..123A},
\texttt{emcee} \citep{Foreman_Mackey:2013},
\texttt{fitsio},\footnote{\url{https://github.com/esheldon/fitsio}}
\texttt{healpix} \citep{Gorski:2005},\footnote{\url{http://healpix.sourceforge.net}}
\texttt{healpy} \citep{Zonca:2019},\footnote{\url{https://github.com/healpy/healpy}}
\texttt{healsparse},\footnote{\url{https://healsparse.readthedocs.io/en/latest/}}
\texttt{matplotlib} \citep{Hunter:2007},
\texttt{numpy} \citep{NumPy:2020},
\texttt{simple} \citep{Bechtol:2015},
\texttt{scipy} \citep{Scipy:2020},
\texttt{skymap},\footnote{\url{https://github.com/kadrlica/skymap}}
\texttt{ugali} \citep{Bechtol:2015, MWCensus1},
}


\bibliography{main}{}
\bibliographystyle{aasjournal}

\appendix
\section{Members stars of Carina IV}
\label{appendix:carina4_stars}

In Section~\ref{sec:followup}, we discuss the \textit{Gaia} proper motions and CaHK photometric metallicity measurements of Carina~IV candidate member stars, which further confirm its identity as a metal-poor Milky Way satellite. To facilitate spectroscopic follow-up of Carina~IV member stars, we include in Table~\ref{table:Carina4stars} the measured proper motions and metallicities for all stars with $g_0 < 21.3$ and high \texttt{ugali} membership probabilities \cytwo{($p_{\rm ugali} > 0.2$). For the other two systems, Phoenix III and DELVE 7, as shown in Fig. \ref{fig:ugali}, we find very few stars with high \texttt{ugali} membership probabilities above the Gaia detection limit of $G \sim 21$. For Phoenix III, the only source is the BHB candidate  \textit{Gaia} DR3 4959457079527646080, with $p_{\rm ugali} = 0.99$, $g_0 = 20.8$, and $r_0 = 21.0$. This star however does not have a \textit{Gaia} proper-motion measurement.  In the case of DELVE 7, none of the high-probability member stars are detected by Gaia.}

\FloatBarrier

\begin{deluxetable*}{cccccccccc}
\label{table:keck_members}
\tablecaption{\label{table:Carina4stars}Properties of stars with high \texttt{ugali} membership probabilities for Carina IV ($p_{\rm ugali} > 0.2$) and $g_0 < 21.3$. The stars are divided into three categories: (1) likely members, characterized by consistent \textit{Gaia} proper motion measurements and low CaHK metallicities (${\rm [Fe/H]_{\rm CaHK}} < -1.5$); (2) possible members, with consistent \textit{Gaia} proper motions but higher metallicities; and (3) non-members, exhibiting inconsistent proper motions. }
\tablehead{Star Name & RA & DEC & $g_0$ & $r_0$ & $\mu_{\alpha*}$ & $\mu_\delta$ &[Fe/H]$_{\rm CaHK}$ & $p_{\rm ugali}$  & Type \\
     & (deg) & (deg) & (mag)&  (mag) & (km s$^{-1}$) & (km s$^{-1}$)  & (dex)  &  & }
\startdata
GAIA DR3 5286348758412769920 & 104.085 & -63.314 & 20.1 & 19.5 & 0.8 $\pm$ 0.6 & 0.1 $\pm$ 0.5 & -2.4 $\pm$ 0.3 & 0.45  & RGB \\
GAIA DR3 5286348934506327424 & 104.031 & -63.306 & 20.4 & 20.4 & 1.1 $\pm$ 1.4 & -0.3 $\pm$ 1.2 &- & 0.96 & BHB  \\
GAIA DR3 5286348865786852864 & 104.057 & -63.298 & 20.5 & 20.6 & 1.6 $\pm$ 3.0 & 4.0 $\pm$ 2.2 & - & 0.96  & BHB \\
GAIA DR3 5286348865787318784 & 104.032 & -63.308 & 21.1 & 20.6 & 0.9 $\pm$ 2.7 & 0.6 $\pm$ 2.6 & -1.5 $\pm$ 0.3 & 0.89    & RGB \\
GAIA DR3 5286348693988641024 & 104.082 & -63.324 & 21.2 & 20.6 & - & - & -1.6 $\pm$ 0.3 & 0.82   & RGB \\
GAIA DR3 5286348827132240768 & 104.016 & -63.319 & 21.2 & 20.7 & - & - & -2.6 $\pm$ 0.6 & 0.90  & RGB \\
\hline
GAIA DR3 5286301616851702528 & 104.039 & -63.347 & 20.4 & 19.8 & 1.0 $\pm$ 0.7 & 1.4 $\pm$ 0.6 & -1.2 $\pm$ 0.3 & 0.36 & RGB(?)  \\
GAIA DR3 5286348968866071168 & 104.082 & -63.294 & 20.8 & 20.3 & 2.1 $\pm$ 1.2 & -0.2 $\pm$ 0.9 & -1.2 $\pm$ 0.4 & 0.39  & RGB(?)  \\
GAIA DR3 5286301655509365888 & 104.063 & -63.345 & 20.9 & 20.4 & 0.5 $\pm$ 1.6 & -0.3 $\pm$ 1.2 & -0.6 $\pm$ 0.4 & 0.61 & RGB(?) \\
\hline
GAIA DR3 5286348865786859264 & 104.052 & -63.313 & 20.7 & 20.1 & 3.6 $\pm$ 1.1 & 6.6 $\pm$ 0.8 & -0.6 $\pm$ 0.4 & - & Non-member  \\
GAIA DR3 5286301513772625536 & 104.154 & -63.335 & 20.9 & 20.3 & 4.8 $\pm$ 1.4 & 5.1 $\pm$ 0.9 & -1.4 $\pm$ 0.3 &  - & Non-member  \\
\enddata
\end{deluxetable*}


\onecolumngrid

\newpage
\section{\cytwo{Phoenix III as a possible intermediate age Star Cluster}}
\label{appendix:phx3sc}
\begin{figure}[H]
    \centering
    \includegraphics[width=\linewidth]{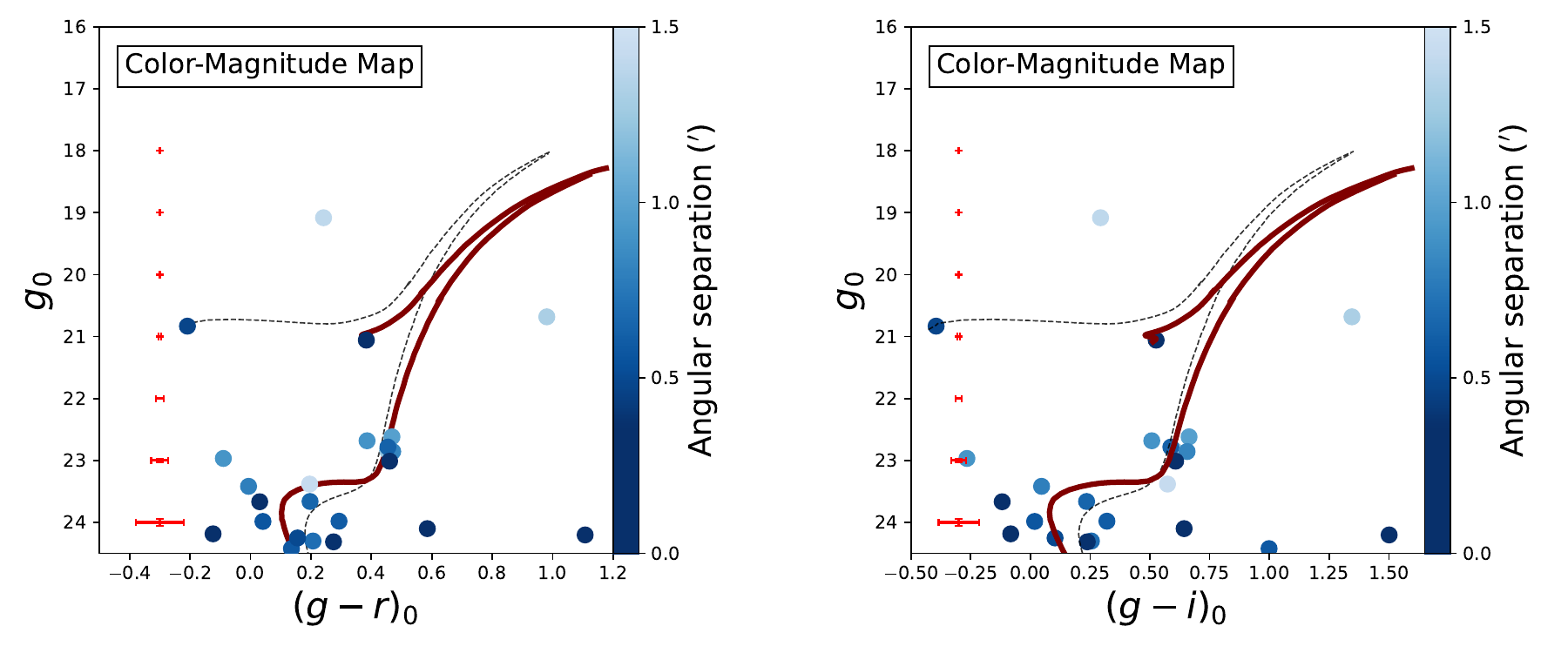}
    \caption{\label{fig:alt_phx3} \cytwo{ Color--magnitude diagrams for both $g-r$ (left) and $g-i$ (right) band pairs for stars within $1\farcm5$ of Phoenix III.  Stars are colored by separation from each system's centroid, with darker colors representing smaller separations. We overlay \texttt{PARSEC} isochrones corresponding to the intermediate-age 7~Gyr solution (solid maroon lines) and the old 12~Gyr solution (black dotted lines) for comparison. The median photometric uncertainties for stars around Phoenix III as a function of $g$-band magnitude, are indicated by red error bars.}}
\end{figure}

\cytwo{As discussed in Section~\ref{sec:phoenixIII}, the morphological fits for Phoenix~III yield two viable solutions. One corresponds to an old, metal-poor stellar population with an age of approximately 12~Gyr, which we argue is the more likely interpretation based on its inclusion of the candidate BHB star \textit{Gaia} DR3 4959457079527646080. Upcoming HST follow-up observations are expected to confirm or reject this assumption \citep{McQuinn:2025hst..prop18066M}. However, for completeness, we present in the appendix a brief description of the intermediate-age ($\sim$7~Gyr) alternative solution identified by \texttt{ugali}. For this alternative solution, the posterior peaks at a distance modulus of of $(m-M)_0 = 20.5_{-0.1}^{+0.2}$,   age of $\tau = 7.0_{-1.5}^{+1.7}$ and a metallicity of $Z_{\rm phot} = 0.00040_{-0.00016}^{+0.00022}$. In Figure~\ref{fig:alt_phx3}, we compare this intermediate-age solution with the old 12~Gyr solution adopted in the main analysis using color–magnitude diagrams in the $g-r$ band (as in the main analysis) and the additional $g-i$ band information.}

\end{document}